\documentclass[12pt,a4]{article}
\usepackage{amsthm,amsmath,latexsym,multibox,amssymb,amsfonts,amsbsy}
\usepackage{graphicx,lscape,fancyhdr,array,stmaryrd,euscript,wrapfig}
\usepackage{cite}
\usepackage{amsthm,amsmath,latexsym,multibox,amssymb,amsfonts}
\usepackage{graphicx,lscape,fancyhdr,array,stmaryrd}

\newcommand{\bep}{\begin{picture}}
\newcommand{\eep}{\end{picture}}

\newcounter{YoungHeight}\newcounter{YoungWidth}

\newcounter{Mul1}\newcounter{Mul2}\newcounter{Mul3}\newcounter{Mul4}
\newcounter{A0}\newcounter{A1}\newcounter{A2}\newcounter{A3}\newcounter{A4}\newcounter{A5}\newcounter{A6}
\newcounter{B0}\newcounter{B1}\newcounter{B2}\newcounter{B3}
\newcounter{C1}\newcounter{C2}\newcounter{C3}\newcounter{C4}\newcounter{C6}\newcounter{C7}
\newcounter{D1}\newcounter{D2}\newcounter{D3}\newcounter{D4}\newcounter{D5}
\newcounter{T0}\newcounter{T1}

\newcounter{TGR0}
\newcounter{R0}\newcounter{R1}\newcounter{R2}\newcounter{R3}
\newcounter{AR0}\newcounter{AR1}\newcounter{AR2}\newcounter{AR3}\newcounter{AR4}\newcounter{AR5}\newcounter{AR6}\newcounter{AR7}
\newcounter{Dotted0}\newcounter{Dotted1}\newcounter{Dotted2}\newcounter{Dotted3}

\newcounter{reptA}

\newlength{\txtHShift}

\newlength{\txtWidth}

\newcommand{\HalfLength}[2]{\setcounter{Mul1}{#1}\setcounter{Mul2}{#1}\addtocounter{Mul1}{\value{Mul2}}\addtocounter{Mul1}{\value{Mul2}}%
\addtocounter{Mul1}{\value{Mul2}}\addtocounter{Mul1}{\value{Mul2}}\setcounter{#2}{\value{Mul1}}}

\newcommand{\Add}[3]{\setcounter{#1}{#2}\addtocounter{#1}{#3}}

\newcommand{\Length}[1]{#10}
\newcommand{\YoungScale}{}

\newcommand{\shiftedText}[2]{{\hspace{#1}#2}}
\newcommand{\calcHShift}[1]{\settowidth{\txtWidth}{#1}\setlength{\txtHShift}{-0.5\txtWidth}}

\newcommand{\TextCenterB}[3]{{\calcHShift{#1}\HalfLength{#2}{T0}\Add{T1}{\Length{#3}}{-7}\put(\value{T0},\value{T1}){\shiftedText{\txtHShift}{#1}}}}

\newcommand{\TextTop}[3]{{\calcHShift{#1}\HalfLength{#2}{T0}\Add{T1}{\Length{#3}}{-9}\put(\value{T0},\value{T1}){\shiftedText{\txtHShift}{#1}}}}

\newcommand{\BlockA}[2]{{\YoungScale\bep(\Length{#1},\Length{#2}){\Add{A1}{#1}{1}\Add{A2}{#2}{1}}%
\multiput(0,0)(10,0){\value{A1}}{\line(0,1){\Length{#2}}}\multiput(0,0)(0,10){\value{A2}}{\line(1,0){\Length{#1}}}%
\setcounter{YoungHeight}{\Length{#2}}\setcounter{YoungWidth}{\Length{#1}}\eep}}

\newcommand{\BlockB}[4]{{\YoungScale\Add{B3}{\Length{#2}}{\Length{#4}}%
\bep(\Length{#1},\value{B3})\put(0,\Length{#4}){\BlockA{#1}{#2}}%
\put(0,0){\BlockA{#3}{#4}}\setcounter{YoungHeight}{\value{B3}}\setcounter{YoungWidth}{\Length{#1}}\eep}}

\newcommand{\RectT}[3]{\bep(\Length{#1},\Length{#2})\put(0,0){\line(1,0){\Length{#1}}}\put(0,0){\line(0,1){\Length{#2}}}%
\put(\Length{#1},\Length{#2}){\line(-1,0){\Length{#1}}}\put(\Length{#1},\Length{#2}){\line(0,-1){\Length{#2}}}#3{#1}{#2}\eep}

\newcommand{\RectARow}[2]{{\bep(\Length{#1},10)\put(0,0){\RectT{#1}{1}{\TextTop{#2}}}\eep}}

\newcommand{\RectBRow}[4]{{\bep(\Length{#1},20)\put(0,0){\RectT{#2}{1}{\TextTop{#4}}}%
\put(0,10){\RectT{#1}{1}{\TextTop{#3}}}\eep}}

\newcommand{\RectARowUp}[2]{{\bep(\Length{#1},10)\put(0,0){\RectT{#1}{1}{\TextCenterB{#2}}}\eep}}

\newcommand{\RectCRowUp}[6]{{\bep(\Length{#1},30)\put(0,0){\RectT{#3}{1}{\TextCenterB{#6}}}%
\put(0,10){\RectT{#2}{1}{\TextCenterB{#5}}}\put(0,20){\RectT{#1}{1}{\TextCenterB{#4}}}\eep}}

\newcommand{\RectBYoung}[3]{{\bep(0,0)\put(0,0){#3}\eep\Add{A0}{\value{YoungHeight}}{10}%
\bep(\Length{#1},\value{A0})%
\put(0,\value{YoungHeight}){\RectT{#1}{1}{\TextTop{#2}}}\eep}}

\newcommand{\YoungA}{\BlockA{1}{1}}
\newcommand{\YoungB}{\BlockA{2}{1}}

\newcommand{\YoungAA}{\BlockA{1}{2}}
\newcommand{\YoungBA}{\BlockB{2}{1}{1}{1}}
\newcommand{\YoungBB}{\BlockA{2}{2}}

\newcommand{\YoungBAA}{\BlockB{2}{1}{1}{2}}

\newcommand{\BlockBoldA}[2]{{\YoungScale\bep(\Length{#1},\Length{#2}){\linethickness{2mm}\Add{A1}{#1}{1}\Add{A2}{#2}{1}}%
\multiput(0,0)(10,0){\value{A1}}{\linethickness{0.4mm}\line(0,1){\Length{#2}}}\multiput(0,0)(0,10){\value{A2}}{\linethickness{0.4mm}\line(1,0){\Length{#1}}}\eep}}
\newcommand{\BlockBoldB}[4]{{\YoungScale\Add{B3}{\Length{#2}}{\Length{#4}}\bep(\Length{#1},\value{B3})\put(0,\Length{#4}){\BlockBoldA{#1}{#2}}%
\put(0,0){\BlockBoldA{#3}{#4}}\eep}}

\newcommand{\YoungBoldA}{\BlockBoldA{1}{1}}

\newcommand{\YoungBoldBAA}{\BlockBoldB{2}{1}{1}{2}}

\newcommand{\stackTB}[2]{{\begin{tabular}{|c|}\hline #1\\ #2\\ \hline\end{tabular}}}

\newtheorem*{Lemma}{Lemma}

\newcommand{\Y}[1]{{\ensuremath{\mathbb{Y}\{#1\}}}}
\newcommand{\Yy}{\ensuremath{\mathbf{Y}}}
\newcommand{\Xx}{\ensuremath{\mathbf{X}}}
\newcommand{\Zz}{\ensuremath{\mathbf{Z}}}
\newcommand{\Ss}{\ensuremath{\mathbf{S}}}
\newcommand{\Ll}{\ensuremath{\mathbf{L}}}
\newcommand{\Ds}{\ensuremath{\mathbf{A}}}
\newcommand{\Ww}{\ensuremath{\mathbf{W}}}

\newcommand{\bs}{{\ensuremath{\mathbf{s}}}}

\newcommand{\DL}{{D}}

\newcommand{\DO}{{D_{\Omega}}}
\newcommand{\DOTwisted}{{\widetilde{D}_{\Omega}}}

\newcommand{\Sigp}{{\ensuremath{{\boldsymbol \sigma}_+}}}
\newcommand{\Sigm}{{\ensuremath{{\boldsymbol \sigma}_-}}}

\newcommand{\te}{{\tilde{e}}}
\newcommand{\tomega}{{\tilde{\omega}}}

\newcommand{\ComplexF}[6]{\ensuremath{0 \longrightarrow #1\longrightarrow #2 \longrightarrow #3 \longrightarrow #4  \longrightarrow #5 \longrightarrow #6 \longrightarrow 0}}

\newcommand{\AlgebraFont}[1]{\mathfrak{#1}}

\newcommand{\DSADS}{{\ensuremath{\boldsymbol{(A)dS_d}}}}
\newcommand{\dSAdS}{{\ensuremath{(A)dS_d}}}

\newcommand{\ads}{{\ensuremath{\AlgebraFont{so}(d-1,2)}}}
\newcommand{\ds}{{\ensuremath{\AlgebraFont{so}(d,1)}}}
\newcommand{\iso}{{\ensuremath{\AlgebraFont{iso}(d-1,1)}}}

\newcommand{\lorentz}{{\ensuremath{\AlgebraFont{so}(d-1,1)}}}

\newcommand{\msv}{{\ensuremath{\AlgebraFont{so}(d-1)}}}
\newcommand{\mls}{{\ensuremath{\AlgebraFont{so}(d-2)}}}
\newcommand{\mlscogroup}{{\ensuremath{\AlgebraFont{iso}(d-2)}}}

\newcommand{\sod}{{\ensuremath{so(d)}}}

\newcommand{\sld}{{\ensuremath{sl(d)}}}

\newcommand{\lorentzS}{{\ensuremath{\scriptstyle{\AlgebraFont{so}(d-1,1)}}}}

\newcommand{\Verma}[2]{\ensuremath{\EuScript{D}\left({\textstyle{#1}};#2\right)}}

\newcommand{\Irrep}[2]{\ensuremath{\EuScript{H}\left({\textstyle{#1}};#2\right)}}

\newcommand{\pl}{\partial}

\newcommand{\fm}[1]{_{\boldsymbol{{#1}}}}

\newcommand{\be}{\begin{equation}}
\newcommand{\ee}{\end{equation}}
\newcommand{\bes}{\begin{split}}
\newcommand{\es}{\end{split}}
\newcommand{\bee}{\begin{eqnarray}}
\newcommand{\eee}{\end{eqnarray}}
\newcommand{\beee}{\begin{array}}
\newcommand{\bem}{\begin{multline}}
\newcommand{\eem}{\end{multline}}
\newcommand{\bec}{\begin{Comment}}
\newcommand{\ec}{\end{Comment}}

\newcommand{\Sa}{{\ensuremath{\mathbf{a}}}}

\newcommand{\la}{{\ensuremath{\mathsf{a}}}}

\newcommand{\lc}{{\ensuremath{\mathsf{c}}}}

\newcommand{\aI}{{\ensuremath{\mathcal{I}}}}

\newcommand{\smallpic}[1]{{\unitlength=0.2mm#1}}

\newcommand{\boldpic}[1]{{\linethickness{0.4mm}#1}}
\newcommand{\smallboldpic}[1]{{\unitlength=0.2mm\linethickness{0.35mm}#1}}

\renewcommand{\theequation}{\arabic{section}.\arabic{equation}}

\begin{document}
\renewcommand{\thefootnote}{\fnsymbol{footnote}}
{\begin{titlepage}
\begin{flushright}
\vspace{1mm}
FIAN/TD/05-09\\
\end{flushright}

\vspace{1cm}

\begin{center}
{\bf \Large Gauge Fields in $\DSADS$ and Connections of its symmetry algebra } \vspace{1cm}

\textsc{E.D. Skvortsov\footnote{skvortsov@lpi.ru}}

\vspace{.7cm}

{ I.E.Tamm Department of Theoretical Physics, P.N.Lebedev Institute of Physics,\\Leninsky prospect 53, 119991, Moscow, Russia}
\end{center}

\vspace{0.5cm}

\begin{abstract}
The generalized connections of the de Sitter algebra \ds{} and anti-de Sitter algebra \ads, which are differential forms of
arbitrary degree with values in any irreducible (spin)-tensor representation of the (anti)-de Sitter algebra, are studied. It is
shown that arbitrary-spin gauge field in (anti)-de Sitter space, massless or partially-massless, can be described by a single
connection. A 'one-to-one' correspondence between the connections of the (anti)-de Sitter algebra and the gauge fields is
established. The gauge symmetry is manifest and auxiliary fields are automatically included in the formalism.
\end{abstract}
\end{titlepage}
\renewcommand{\thefootnote}{\arabic{footnote}}
\setcounter{footnote}{0}

\section*{Introduction and Main Results}\setcounter{equation}{0}

The paper aims at (\textbf{I}) studying generalized Yang-Mills connections of the (anti)-de Sitter algebra, i.e. \ds{} (de
Sitter) or \ads{} (anti-de Sitter), that are defined to be differential forms of arbitrary degree with values in any irreducible
finite-dimensional representation of the (anti)-de Sitter algebra; (\textbf{II}) constructing the frame-like formulation for
gauge fields in (anti)-de Sitter space, including all types of massless and partially-massless fields.

Our motivation is twofold: first, to study a natural geometric and algebraic object presented by a generalized connection of the
space-time symmetry algebra, the generalization lies in allowing the form degree and the representation in which a generalized
connection takes values to be arbitrary, with the Yang-Mills connection arising if the form degree is $1$ and the representation
is the adjoint one. Second, to develop the theory involving the fields of the most general spin type in higher-dimensions, the
higher-spin theory.

For many years the theory of higher-spin gauge fields, which studies the classical problem of constructing consistent interacting
theories of fields of various spins, has attracted considerable interest. One of the main goals of higher-spin theory is the full
classical nonlinear theory of massless fields of spins $s=0,1,2,...$ constructed in \cite{Vasiliev:1990en, Vasiliev:2003ev},
introducing two new ingredients - the unfolded approach to field equations \cite{Vasiliev:1988xc, Vasiliev:1988sa,
Vasiliev:1992gr}, which is based on free differential algebras \cite{Sullivan77,Nieuwenhuizen:1982zf, D'Auria:1982pm,
D'Auria:1982nx} and the higher-spin algebras \cite{Konstein:1989ij, Konshtein:1988yg, Vasiliev:2004cm}, which are certain
infinite-dimensional extensions of the space-time symmetry algebra. For recent reviews on higher-spin gauge theory see
\cite{Buchbinder:2001bs,Francia:2002aa,Sezgin:2002rt,Bekaert:2003uc,Sagnotti:2003qa,Sorokin:2004ie,
Bouatta:2004kk,Bekaert:2005vh,Francia:2005bv,Sagnotti:2005ns}.

In $d=4$ the spin degrees of freedom are determined by a single (half)integer $s=0,\frac12,1,...$ . Beyond $d=4$, more general
type of fields come into play, whose both the spin (physical polarization tensor) and the field potential are neither symmetric
nor antisymmetric tensors \cite{Curtright:1980yk, Labastida:1987kw, Zinoviev:2002ye, Bekaert:2002dt, Zinoviev:2003dd,
Zinoviev:2003ix, Alkalaev:2003hc, Alkalaev:2006hq, Bekaert:2006ix, Moshin:2007jt, Skvortsov:2008sh, Reshetnyak:2008gp,
Skvortsov:2008vs, Campoleoni:2008jq, Zinoviev:2008ve, Alkalaev:2008gi,Zinoviev:2009vy}. These fields of a general tensor type are
referred to as mixed-symmetry fields and are more difficult to study even at the free level.

The major motivation for studying the gauge fields rather than massive ones is that the gauge symmetry is very restrictive. For
the case of spin-$s$ fields, the gauge symmetry is known to provide a very limited class of higher-spin multiplets
\cite{Konstein:1989ij, Konshtein:1988yg, Vasiliev:2004cm}, each containing fields of arbitrary large spins, and to fix all
dimensionless coupling constants for the vertices of spin-$s$ fields \cite{Vasiliev:2003ev}.

The massive modes of string theory are believed to come via spontaneous breaking of higher-spin gauge symmetries
\cite{Gross:1988ue,Sundborg:2000wp,Sezgin:2001zs,Klebanov:2002ja}. Massless higher-spin fields are also known to appear in the
tensionless limit of string theory \cite{Francia:2002pt, Sagnotti:2003qa, Bonelli:2003kh, Francia:2005bu, Francia:2006hp}.

Of most interest is the theory of arbitrary-spin fields, in particular of mixed-symmetry fields, in the (anti)-de Sitter
background. In the Minkowski space the only gauge fields are massless ones. A wider variety of gauge fields is available in the
(anti)-de Sitter space. The gauge fields in (anti)-de Sitter space are presented by different types of massless
\cite{Nicolai:1984hb, Metsaev:1995re, Metsaev:1997nj, Brink:2000ag} and partially-massless fields \cite{Deser:1983mm,
Deser:1983tm, Higuchi:1986wu, Deser:2001pe, Deser:2001xr, Deser:2001us, Deser:2003gw, Zinoviev:2001dt}, whose geometric and
manifestly gauge invariant description in terms of the generalized connections will be constructed below.

(\textbf{I}) An $\dSAdS$ gauge connection $W^\Ds\fm{q}$ is defined by a pair $\{\mathbf{q},\Ds\}$, where $q=1,...,d$ is a form
degree and $\Ds$ is a finite-dimensional irreducible representation of the (anti)-de Sitter algebra, i.e. either a tensor or a
spin-tensor, which is convenient to specify by a Young diagram\footnote{In this paper, 'an irreducible tensor of \sld{} or \sod{}
having the symmetry of Young diagram $\Y{s_1,...,s_n}$' is synonymous to 'a finite-dimensional irreducible highest weight module
of \sld{} or \sod{} with highest weight $(s_1,...,s_n,0_{n+1},...,0_\nu)$', where $\nu=d-1$ for \sld{} and $\nu=[d/2]$ for \sod.
To avoid (anti)-selfdual representations in the \sod{} case we assume $\nu>n$. 'A tensor having the symmetry of $\Yy$' refers
only to the permutation symmetry of its indices, which for $\sod$ is weaker than irreducibility.}. Given a flat $\dSAdS$
covariant derivative $\DO$, i.e. $\DO^2=0$, the field strength $R^\Ds\fm{q+1}=\DO W^\Ds\fm{q}$ is manifestly invariant under the
gauge transformations $\delta W^\Ds\fm{q}=\DO \xi^\Ds\fm{q-1}$, providing us with a natural framework for gauge theories in
$\dSAdS$ \cite{Vasiliev:2001wa, Alkalaev:2003qv, Skvortsov:2006at}.

(\textbf{II}) A gauge field in (anti)-de Sitter space is defined \cite{Alkalaev:2009} by a triple $(\Ss,q,t)$, where $\Ss$ is a
Young diagram that specifies both the symmetry type of the field potential $\phi^\Ss(x)$ as a Lorentz tensor and the physical
polarization tensor of \msv{}, which is the Wigner little algebra of \dSAdS{} \cite{Brink:2000ag}. The integers $q$ and $t$
determine the tensor type $\Ss_1$ of the gauge parameter $\xi^{\Ss_1}(x)$ and the gauge transformation law. Let
$\Ss=\Y{s_1,...,s_p}$, then the gauge parameter $\xi^{\Ss_1}(x)$ is a Lorentz tensor having the symmetry of
$\Ss_1=\Y{s_1,...,s_{q-1},s_q-t,s_{q+1},...,s_p}$. $t$ is equal to the order of derivative in the gauge transformations
\begin{align*} \delta\phi^{\Ss}=\overbrace{D...D}^t \xi^{\Ss_1}+...\quad .\end{align*} The irreducible representation of the
$\dSAdS$-algebra is realized on the solutions of certain gauge invariant equations imposed on $\phi^{\Ss}$.

(\textbf{I vs. II}) In this paper we address the question: given a pair $\{\mathbf{q},\Ds\}$, what type of $\dSAdS$ field does
the gauge connection $W^\Ds\fm{q}$ describe? Does this map cover the whole variety of $\dSAdS$ fields? We will see the answer on
the latter question is yes, i.e. to every given triple $(\Ss,q,t)$ one can assign certain gauge connection $W^\Ds\fm{q}$.

To be precise, the main result is that a gauge field defined by $(\Ss,q,t)$ can be described by a single degree-$q$ differential
form $W^\Ds\fm{q}$ over (anti)-de Sitter space with values in an irreducible tensor representation of the (anti)-de Sitter
algebra that is characterized by Young diagram $\Ds=\Y{s_1-1,...,s_{q}-1,s_q-t,s_{q+1},...,s_p}$.
\begin{align}
&\Ss=\parbox{3.6cm}{{\bep(80,80)\unitlength=0.4mm%
\put(0,10){\RectCRowUp{5}{4}{3}{$s_{q+1}$}{$...$}{$s_p$}}%
\put(0,40){\RectCRowUp{10}{9}{8}{$s_1$}{$...$}{$s_q$}}\put(60,40){\YoungA}%
\put(0,40){\YoungBAA}\put(80,50){\YoungBA}\put(70,40){\YoungAA}%
\put(68,64){$...$}\put(24,64){$...$}%
\eep}},&&\Ds=\parbox{3.6cm}{\boldpic{\bep(80,80)\unitlength=0.4mm%
\put(0,0){\RectCRowUp{5}{4}{3}{$s_{q+1}$}{$...$}{$s_p$}}%
\put(0,30){\RectCRowUp{8}{7}{6}{$...$}{$s_q-1$}{$s_q-t$}}\put(60,40){\YoungBoldA}%
\put(0,40){\YoungBoldBAA}\put(80,60){\YoungBoldA}\put(70,50){\YoungBoldA}%
\put(68,64){$...$}\put(24,64){$...$}\put(0,60){\RectARowUp{9}{$\quad s_1-1$}}%
\eep}} \nonumber\end{align} Decomposing $\dSAdS$-module $\Ds$ with respect to the Lorentz subalgebra of the (anti)-de Sitter
algebra, $W^\Ds\fm{q}$ yields a collection of Lorentz connections. The dynamical field $\phi^{a(s_1),...,u(s_p)}$ is embedded
into the generalized frame field, with the rest of connections playing auxiliary role at the free level. Setting certain
components of the field strength $R^\Ds\fm{q+1}$ to zero, the correct equations on $\phi^{\Ss}$ are obtained.

The approach is a far-reaching generalization of the MacDowell-Mansouri-Stelle-West approach to gravity \cite{MacDowell:1977jt,
Stelle:1979aj}, in which the single connection\footnote{Indices $A,B,C,...=0,...,d$ are either of the de Sitter \ds{} or anti-de
Sitter  \ads{} algebra. Indices $a,b,c,...=0,...,d-1$ are of the fiber Lorentz algebra \lorentz. Indices $\mu,\nu,...=0,...,d-1$
are world indices of differential forms.} $W^{A,B}_\mu dx^\mu=-W^{B,A}_\mu dx^\mu$ of the (anti)-de Sitter algebra, after
breaking the (anti)-de Sitter algebra down to the Lorentz algebra yields the frame (vielbein/tetrad) field $e^a_\mu dx^\mu$ and
the Lorentz spin-connection $\omega^{a,b}_\mu dx^\mu$. It is also a direct extension of the works \cite{Vasiliev:2001wa,
Alkalaev:2003qv, Skvortsov:2006at}, where certain gauge connections of the (anti)-de Sitter algebra were proposed as a natural
framework for the $(\Y{s},1,1)$, $(\Y{s_1,...,s_p},1,1)$\footnote{To be precise, in \cite{Alkalaev:2003qv} $q$ is the number of
first equal rows of $\Ss$.} and $(\Y{s},1,t)$ series of gauge fields.

The paper organized as follows. In Section \ref{SWaveEquations} we discuss wave equations in Minkowski, de Sitter and anti-de
Sitter spaces and their relation to the representation theory. The precise definition and classification of fields in (anti)-de
Sitter space is given in Section \ref{SGaugeFieldsdSAdS}. In Section \ref{SBackgroundGeometry} we recall the description of the
(anti)-de Sitter geometry by Cartan connections. The main subject of the paper, gauge connections of the $\dSAdS$-algebra, is
studied in Section \ref{SGaugeConnections}. The correspondence between gauge fields in $\dSAdS$ and gauge connections of the
(anti)-de Sitter algebra is established in Section \ref{SGeneralCase}. The discussion of the results and further developments
concerning the nonlinear theory of gauge fields are in Section \ref{SConclusions}.

In the next section we review without details the background for field theories in Minkowski and (anti)-de Sitter spaces,
accentuating the difference between them. Then, we argue that the frame-like approach and its generalization to arbitrary-spin
fields are more powerful ones and illustrate on the example of a massless spin-$s$ field the advantage of describing fields by a
single gauge connection.

\section*{Field theories in Minkowski and $\DSADS$, Mixed-Symmetry Fields}\setcounter{equation}{0}
\renewcommand{\theequation}{\arabic{equation}}
\setcounter{section}{0} \setcounter{equation}{0}

Relativistic fields are known to be in one-to-one correspondence with unitary irreducible representations of the space symmetry
algebra, being \iso\ for a $d$-dimensional Minkowski space. The famous Wigner results \cite{Wigner:1939cj} on the classification
of relativistic fields in $4d$ Minkowski spacetime can be straightforwardly generalized to arbitrary spacetime dimension $d\geq4$
\cite{Bekaert:2006py}.

As in $4d$, a unitary irreducible representation of \iso\ is determined by the two parameters, the mass $m^2\geq0$ and the spin
$\Ss$. The mass fixes the Casimir $P_aP^a$ of \iso{}. The spin defines an irreducible representation of the stability algebra
$\mathfrak{f}$ of the momentum that obeys $P_aP^a=m^2$. Called the Wigner little algebra, $\mathfrak{f}$ is either \msv{} for
time-like momentum ($m^2>0$) or \mlscogroup{} for light-like momentum ($m^2=0$). Due to the requirement for the number of spin
degrees of freedom to be finite the translations of \mlscogroup{} must be represented trivially, reducing $\mathfrak{f}$ to
\mls{} for $m^2=0$. Therefore, the spin degrees of freedom are in one-to-one correspondence with irreducible (spin)-tensor
representations of \msv{} or \mls.

Having completed the classification of relativistic fields, the next problem, referred to as the Bargmann-Wigner program
\cite{Gelfand:1948, Bargmann:1948ck}, is to associate with each pair $(m^2,\Ss)$ a relativistic wave equation whose solution
space forms the representation of \iso{} labelled by $m^2$ and $\Ss$. The wave equation has the form
$(\square+m^2)\phi^{ab...}(x)=0$ with $\phi^{ab...}(x)$ being certain (spin)-tensor field of the Lorentz algebra and may be
supplemented with some algebraic and differential constraints imposed on $\phi^{ab...}$ to exclude the spin states different from
$\Ss$.

If there are no additional requirements to be met, e.g., that the wave equation together with the differential constraints come
from a Lagrangian, without loss of generality $\phi^{ab...}$ can take values in an irreducible (spin)-tensor representation of
the Lorentz algebra, say in $\mathbf{R}$. Given $m^2$ and $\Ss$ there exist infinitely many choices of $\mathbf{R}$, known as
dual descriptions. Despite this ambiguity, natural is to take $\mathbf{R}$ to have the same symmetry properties as the physical
polarization tensor (spin) has, i.e., to take $\mathbf{R}=\Ss$ as Young diagrams. For this remarkable choice $\phi^\Ss(x)$ will
be called a spin-$\Ss$ potential. Representing fields by potentials appears to be more fundamental since, for example,
electro-magnetic interactions requires potential $A_\mu$ rather than the Faraday tensor $F_{\mu\nu}$.

In 4d the spin $\Ss$ is defined by a single (half)-integer, say $\bs$, which corresponds to a totally-symmetric rank-$\bs$
(spin)-tensor potetial. In higher dimensions there exist more complicated (spin)-tensor representations of the Wigner little
algebra, referred to as mixed-symmetry (spin)-tensors, which are neither symmetric nor antisymmetric (spin)-tensors. This being
the case, the spin $\Ss$ is defined by a number of (half)-integers $\bs_1$,...,$\bs_p$. The maximal value of $p$ is equal to
$[(d-1)/2]$ for massive fields and to $[(d-2)/2]$ for massless ones.

Genuine massless mixed-symmetry fields \cite{Curtright:1980yk, Labastida:1987kw, Zinoviev:2002ye, Bekaert:2002dt,
Zinoviev:2003dd, Zinoviev:2003ix, Alkalaev:2003hc, Alkalaev:2006hq, Bekaert:2006ix, Moshin:2007jt, Skvortsov:2008sh,
Reshetnyak:2008gp, Skvortsov:2008vs, Campoleoni:2008jq, Zinoviev:2008ve, Alkalaev:2008gi, Zinoviev:2009vy}, i.e., those having at
least two different nonzero weights $\bs_1\neq\bs_2\neq0$, have two distinctive features in Minkowski space (i) there are more
than one gauge parameter (gauge parameters are counted by the symmetry type); (ii) the gauge symmetry is reducible, meaning that
one can transform the gauge parameter $\xi^1$ as $\delta \xi^1=\pl\xi^2$ so that $\delta \phi=\pl\xi^1\equiv0$ for such $\xi^1$;
$\xi^1$ and $\xi^2$ are referred to as the first and second level gauge parameters, respectively. There can be arbitrary many
levels in general. For massless fields in Minkowski space there are as many levels as the number of nonzero weights is in \Ss.

In what follows we restrict ourselves mainly to the gauge fields, which are presented in Minkowski space by massless fields and,
as we will see, there are more different types of gauge fields in (anti)-de Sitter space. Massless or, more generally, gauge
fields seem to be more fundamental than massive ones.

The absence of an effective mechanism to control physical degrees of freedom complicates the study of massive fields, even at the
linear level \cite{Singh:1974qz}. The constructive idea, first realized for spin-$\bs$ fields in \cite{Zinoviev:2001dt}, is to
reformulate massive fields as gauge theories with Stueckelberg gauge symmetries. The Lagrangians \cite{Zinoviev:2001dt,
Zinoviev:2002ye, Zinoviev:2003dd, Zinoviev:2003ix} of massive fields are the sums of Lagrangians of massless fields coupled
together via low derivative terms. The number of physical degrees of freedom can be easily controlled at the nonlinear level by
requiring the vertices to be gauge invariant \cite{Zinoviev:2006im, Metsaev:2006ui, Zinoviev:2008ck}. Despite technical problems,
there is no doubt that the approach can be generalized to the fields of any spin, \cite{Zinoviev:2008ve}. There should also be a
yet unknown Higgs-type mechanism allowing to produce massive fields by breaking the higher-spin symmetries of massless fields.

To deform Minkowski mixed-symmetry gauge fields to (anti)-de Sitter space turned out to be a highly nontrivial problem
\cite{Metsaev:1995re, Brink:2000ag, Metsaev:1997nj}, having revealed a great deal of peculiar properties. Only massive fields can
be deformed to $\dSAdS$ without any obstructions.

First, the cosmological constant plays the role of the mass parameter in field equations. Therefore, the massless field ought to
be associated not with the one satisfying $\square \phi=0$ but the one with the wave equation $(\square +..\lambda^2)\phi=0$ that
has a proper gauge invariance, which guarantees the correct number of degrees of freedom propagating on-mass-shell. The gauge
invariance appears generally for nonzero coefficient in front of $\lambda^2$.

Second, the $\dSAdS$ 'Wigner little algebra' is $\msv$ both for massless and massive fields \cite{Metsaev:1995re, Brink:2000ag,
Metsaev:1997nj}. Therefore, it is not possible for the wave equation to be invariant under all gauge symmetries coming from
Minkowski space whatever the mass-like coefficient in front of $\lambda^2$ is. It is the commutativity of translations of
$\mlscogroup$, which is the massless Wigner little algebra in Minkowski space, that allows of multiple gauge symmetries of
mixed-symmetry fields in Minkowski space. It turns out that only one (but any) of the Minkowski gauge symmetries can be
maintained in $\dSAdS$. Because of having less gauge symmetries a gauge field in $\dSAdS$ has more degrees of freedom than the
Minkowski massless field with the same spin. Therefore, it is not possible to deform a generic massless field into $\dSAdS$
smoothly, i.e. without discontinuity in the number of physical degrees of freedom \cite{Brink:2000ag}.

Third feature of $\dSAdS$, discovered for a spin-two field in \cite{Deser:1983mm, Deser:1983tm, Higuchi:1986wu, Deser:2001pe,
Deser:2001xr, Deser:2001us, Deser:2003gw, Zinoviev:2001dt}\footnote{The very term 'partially-massless' was introduced in
\cite{Deser:2001pe}.}, is the existence of a new type of fields: partially-massless fields whose gauge transformation law
contains higher derivatives, and hence they have more degrees of freedom than the massless fields. There is no room for
higher-derivative gauge symmetry in Minkowski space since the corresponding \iso-module realized on the solutions of the wave
equation would be reducible. Due to the noncommutativity of the $\dSAdS$-translations the quotient module becomes irreducible.

The full classification of $\dSAdS$ gauge fields is obtained by collecting different types of massless and partially-massless
fields. $N$ families of gauge fields in $\dSAdS$ are associated with each massless spin-$\Ss$ field in Minkowski space
\cite{Alkalaev:2009}, where $N$ is the number of gauge symmetries in Minkowski space. The first field of each family is called
massless because of the first order gauge transformation law. The fields from the rest of each family contain higher-derivatives
in the gauge transformations and are called partially-massless, the maximal depth of (partially)-masslessness, which counts the
number of derivatives in the gauge transformation law, is determined by the Young diagram $\Ss$.

If field potentials are world tensors, being analogous to the metric field $g_{\mu\nu}$, the approach is referred to as
metric-like. There exists a more powerful approach to gravity in which the gravitational field is represented by a local frame
$e^a_\mu dx^\mu$ and Lorentz spin-connection $\omega^{a,b}_\mu dx^\mu$. The frame-like approach to gravity turned out to be more
fundamental since it is the frame-like approach that allows introducing the gravitational interactions of fermionic fields. For
massless spin-$\bs$ fields the frame-like approach, namely its profound extension known as the unfolded approach
\cite{Vasiliev:1988xc, Vasiliev:1988sa, Vasiliev:1992gr}, turned out to be more fundamental too.

The challenging problem is to construct and classify nonlinear theories involving fields of any spin. The only full classical
nonlinear theory known up to date contains totally-symmetric massless fields \cite{Vasiliev:1989yr, Vasiliev:1990en,
Vasiliev:2003ev}. Its distinguishing features are (i) the theory was constructed within the unfolded approach; (ii) consistent
interactions require nontrivial cosmological constant $\lambda^2\neq0$ \cite{Fradkin:1986qy};  (iii) the underlying symmetry
algebra is certain infinite-dimensional extension of the space-time symmetry algebra satisfying the admissibility condition
\cite{Konshtein:1988yg, Konstein:1989ij, Vasiliev:2004cm}.

The unfolded approach is a reformulation of field equations in terms of free differential algebras \cite{Sullivan77}, which is
the categorial extension of the Lie algebra. The fields within the unfolded approach are differential forms of various degrees
with values in some representations of the space symmetry algebra $\mathfrak{g}$, giving rise upon decomposing with respect to
the Lorentz subalgebra \lorentz\ of $\mathfrak{g}$ to differential forms with fiber Lorentz indices, i.e. to frame-like fields.
The connections $W^\Ds\fm{q}$ proposed for the description of gauge fields in $\dSAdS$ form the gauge module for the
corresponding unfolded system.

In this paper we extend the results of \cite{Vasiliev:2001wa, Alkalaev:2003qv, Skvortsov:2006at} and construct the manifestly
$\dSAdS$-covariant description for arbitrary-spin gauge field in $\dSAdS$ in terms of a single connection of the (anti)-de Sitter
algebra $\mathfrak{g}$, which is \ds{} (de Sitter) or \ads{} (anti-de Sitter). All auxiliary fields turn out to be included in
the connection of $\mathfrak{g}$ automatically. There are two successive reductions of the $\dSAdS$-covariant formulation that
yield the Lorentz-covariant frame-like formulation and, then, the metric-like formulation.

Below, on the example of a massless totally-symmetric field of spin-$s$, we illustrate the evolutionary chain, which is opposite
to the reductions just mentioned, \be\mbox{Lorentz metric-like}\longrightarrow\mbox{Lorentz frame-like}\longrightarrow\dSAdS\
\mbox{connection}.\nonumber\ee

The gauge potential for a totally-symmetric spin-$s$ field is a rank-$s$ symmetric tensor field $\phi_{\mu_1\mu_2...\mu_s}$
subjected to the double-trace constraint \cite{Fronsdal:1978rb} \be\label{FronsdalDoubleTrace}
\eta^{\nu_1\nu_2}\eta^{\nu_3\nu_4}\phi_{\nu_1\nu_2\nu_3\nu_4\mu_5\mu_6...\mu_s}\equiv0.\ee The correct number of physical degrees
of freedom is guaranteed by gauge symmetry \be\delta\phi_{\mu_1...\mu_s}=\DL_{\mu_1}\xi_{\mu_2...\mu_{s}}+\mbox{permutations}\ee
with a rank-$(s-1)$ symmetric traceless gauge parameter $\xi_{\mu_1...\mu_{s-1}}$. It is worth noting that neither the
double-trace constraint nor the gauge invariant equations are self-evident in the metric-like approach, not to mention general
mixed-symmetry fields.

Similar to a spin-two field, spin-$s$ field can also be described within the frame-like approach \cite{Vasiliev:1980as}. The
generalized frame field is a one-form\footnote{A group of $k$ indices in which a tensor is symmetric is denoted by one letter
with the number of indices indicated in round brackets, e.g. $a(k)\equiv a_1a_2...a_k$.} $e^{a(s-1)}_\mu dx^\mu\equiv
e^{a_1...a_{s-1}}_\mu dx^\mu$ that is symmetric and traceless in its $(s-1)$ fiber indices of the Lorentz algebra, i.e., it takes
values in the irreducible representation of \lorentz{} labelled by Young diagram $\smallpic{\RectARow{4}{$\scriptstyle s-1$}}$,
which for $s=2$ reduces to a vector-valued one-form $e^a_\mu dx^\mu$. The linearized gauge transformations read ($h^a_\mu$ is a
background vielbein field) \be \delta e^{a(s-1)}_\mu =\DL_\mu\xi^{a(s-1)}+h^b_\mu\xi^{a(s-1),}_{\phantom{a(s-1),}b}\,, \ee where
the zero-form $\xi^{a(s-1)}$ is a gauge parameter associated with the generalized frame. The shift-symmetry gauge parameter
$\xi^{a(s-1),b}$ represents the generalized local Lorentz transformations, it takes values in the irreducible representation of
\lorentz{} labelled by Young diagram \smallpic{\RectBRow{4}{1}{$\scriptstyle s-1$}{}}. The gauge field associated with
$\xi^{a(s-1),b}$ is a one-form $\omega^{a(s-1),b}_\mu dx^\mu$. The field strength \be
R^{a(s-1)}=De^{a(s-1)}+h^b\wedge\omega^{a(s-1),}_{\phantom{a(s-1),}b}\ee is invariant not only under $\xi^{a(s-1)}$ and
$\xi^{a(s-1),b}$ transformations but under certain algebraic transformations of $\omega^{a(s-1),b}_\mu dx^\mu$ so that the full
gauge law reads \cite{Vasiliev:1980as} \be \delta
\omega^{a(s-1),b}_\mu=\DL_\mu\xi^{a(s-1),b}+h^c_\mu\xi^{a(s-1),b}_{\phantom{a(s-1),b}c}\,, \ee where $\xi^{a(s-1),bb}$ is a
zero-form taking values in the irreducible representation of \lorentz{} labelled by {\smallpic{\RectBYoung{5}{$\scriptstyle
s-1$}{\YoungB}}}. The gauge parameter $\xi^{a(s-1),bb}$ suggests \cite{Lopatin:1987hz} introducing a one-form gauge field
$\omega^{a(s-1),bb}_\mu dx^\mu$ associated with it. The process continues until the gauge field $\omega^{a(s-1),b(s-1)}$ taking
values in \smallpic{\RectBRow{4}{4}{$\scriptstyle s-1$}{$\scriptstyle s-1$}}, so that the full list of the frame-like fields for
a massless spin-$s$ field reads \cite{Lopatin:1987hz}
\begin{align}\label{IntroCollection} e^{a(s-1)}_\mu && \omega^{a(s-1),b}_\mu&& \omega^{a(s-1),bb}_\mu &&...&& \omega^{a(s-1),b(s-2)}_\mu&& \omega^{a(s-1),b(s-1)}_\mu.\end{align}
The fields having more than one index in the second group are called extra inasmuch as these fields are expressed in terms of
higher derivatives of the frame field. The extra fields decouple at the free level, however, they play an important role in the
interacting theory \cite{Vasiliev:1989yr, Vasiliev:1990en, Vasiliev:2001wa, Vasiliev:2003ev}.

In \cite{Vasiliev:2001wa} it was realized that the collection of fields (\ref{IntroCollection}) comes out of a single connection
one-form of the (anti)-de Sitter algebra that takes values in the irreducible representation labelled by a rectangular two-row
Young diagram of length-$(s-1)$, i.e.
\begin{align}\label{IntroDemotionSequence}\parbox{2.3cm}{\boldpic{\RectBRow{6}{6}{$s-1$}{$s-1$}}},\quad W^{A(s-1),B(s-1)}_\mu dx^\mu &&\longrightarrow
&& (\ref{IntroCollection})&&\longrightarrow &&\delta\phi_{\mu(s)}=D_\mu\xi_{\mu(s-1)}.\end{align}

In \cite{Alkalaev:2003qv} the $\dSAdS$-covariant formulation in terms of certain connections of the (anti)-de Sitter algebra was
proposed for a fields of the series $(\Ss,q=1,t=1)$ .

Later, it was recognized in \cite{Skvortsov:2006at} that a partially-massless spin-$s$ field with $t$ derivatives in the gauge
transformation law can be described by a single connection with values in irreducible representation of the (anti)-de Sitter
algebra that has the symmetry of a two-row Young diagram, the lengths of rows being $(s-1)$ and $(s-t)$,
\begin{align}\parbox{2.3cm}{\boldpic{\RectBRow{6}{4}{$s-1$}{$s-t$}}}, \quad W^{A(s-1),B(s-t)}_\mu dx^\mu &&\longrightarrow &&
\delta\phi_{\mu(s)}=\overbrace{D_\mu...D_\mu}^t\xi_{\mu(s-t)}+...\end{align}

Thus, here comes the question, brought up in the introduction, of the correspondence between the gauge fields in $\dSAdS$ and the
connections of the (anti)-de Sitter algebra. In this paper we give the complete answer.

\renewcommand{\theequation}{\arabic{section}.\arabic{equation}}
\section{Wave Equations and Representation theory in Minkowski and \DSADS}\label{SWaveEquations}\setcounter{equation}{0}
Field theory requires (unitary) irreducible representations of the space-time symmetry algebra $\mathfrak{g}$ that are referred
to as massive or massless fields to be realized on the solutions of certain wave equations imposed on tensor fields over the
space-time manifold. Below $\mathfrak{g}$ is \iso{}, \ds{} or \ads.

As it has been already mentioned in the introduction it is most natural to describe a spin-$\Ss$ field by its potential
$\phi^\Ss$ that is a tensor field whose symmetry is determined by $\Ss$ considered as a diagram of the Lorentz algebra. On the
other hand, given a tensor field $\phi^\Ss$ having the symmetry of some Young diagram $\Ss$ it can be referred to as a spin-$\Ss$
field if the proper field equations that single out the physical polarization tensor having the symmetry of $\Ss$ are to be
imposed later on. The physical polarization tensor can be either of $\mls$ or $\msv$ depending on the field type (massless or
massive) and the space-time in question (Minkowski or $\dSAdS$).

Given a mass $m^2$ and a spin $\Ss$, say $\Ss=\Y{s_1,...,s_p}$, let  $\Verma{m^2}{\Ss}$ be a $\mathfrak{g}$-module that is
singled out of the tensor field $\phi^\Ss\equiv\phi^{a(s_1),...,u(s_p)}(x)$ by virtue of\footnote{Recall, that a group of $k$
indices in which a tensor is symmetric is denoted by one letter with the number of indices indicated in round brackets, e.g.
$a(k)\equiv a_1a_2...a_k$. The symmetrization over (groups of) indices denoted by the same letter is implied, e.g.
$c,b(n),c(k)=\frac1{(k+1)!}\sum_{\sigma}c_{\sigma(1)},b(n),c_{\sigma(2)}...c_{\sigma(k+1)}$, which is used to impose Young
conditions in (\ref{FullSystemC}).}
\begin{align}
(\Box+m^2)\phi^{a(s_1),...,u(s_p)}&=0, &&\label{FullSystemA}\\
\DL_m \phi^{a(s_1),...,mc(s_i-1),...,u(s_p)}&=0,&&i=1,...,p,\label{FullSystemB}\\
\phi^{a(s_1),...,b(s_i),...,bc(s_j-1),...,u(s_p)}&\equiv0, &&i,j=1,...,p,\quad i<j,\label{FullSystemC}\\
\eta_{mm}\phi^{a(s_1),...,mmc(s_i-2),...,u(s_p)}&\equiv0, &&i=1,...,p,\label{FullSystemD}\\
\eta_{mm}\phi^{a(s_1),...,mb(s_i-1),...,mc(s_j-1),...,u(s_p)}&\equiv0, &&i,j=1,...,p,\quad i\neq j,\label{FullSystemE}
\end{align}
where $\Box\equiv D_mD^m$ and $D_m$ is the covariant derivative. The constraints fall into two classes: algebraic ones
(\ref{FullSystemC})-(\ref{FullSystemE}), which ensures the algebraic irreducibility of $\phi^\Ss$, i.e. the Young symmetry
(\ref{FullSystemC}) and tracelessness (\ref{FullSystemD})-(\ref{FullSystemE}), and differential ones
(\ref{FullSystemA})-(\ref{FullSystemB}), which put the field on mass-shell (\ref{FullSystemA}) and exclude low spin states
(\ref{FullSystemB}).

The Young symmetry condition (\ref{FullSystemC}) is that the symmetrization of all indices from the $i$-th group of indices with
one index from the $j$-th group provided $i<j$ must vanish. It guarantees that the indices are irreducible under the action of
the permutation group and together with the vanishing trace conditions (\ref{FullSystemD})-(\ref{FullSystemE}) implies that the
tensor is an irreducible Lorentz one.

The Cauchy data are given by one complex function $f^{\la(s_1),...,\lc(s_p)}(p)$ of $(d-1)$ variables that takes values in the
irreducible representation of \msv{} that is characterized by the same Young diagram $\Ss$ as the spin.

An irreducible $\mathfrak{g}$-module that will be referred to as massive or massless spin-$\Ss$ field is denoted by
$\Irrep{m^2}{\Ss}$. Its relation to $\Verma{m^2}{\Ss}$ depends largely on the space-time in question, i.e. on the symmetry
algebra $\mathfrak{g}$, and on the value of the mass parameter $m^2$.

\paragraph{Minkowski space, $\boldsymbol{\mathfrak{g}=\iso}$.}
For the Minkowski case, if $m^2>0$ an irreducible \iso-module $\Irrep{m^2}{\Ss}$ that is referred to as a massive spin-$\Ss$
field with mass $m^2$ is realized on the positive-frequency solutions of (\ref{FullSystemA})-(\ref{FullSystemE}),
$\Verma{m^2}{\Ss}$ is identified with $\Irrep{m^2}{\Ss}$ directly, $\Verma{m^2}{\Ss}=\Irrep{m^2}{\Ss}$.

At $m^2=0$ $\Verma{m^2}{\Ss}$ becomes reducible, signaling the appearance of some gauge symmetry. The gauge symmetry can be
identified with certain modules $\Verma{0}{\Yy^i}$, where $\Yy^i$ determines the symmetry of gauge parameters. For the general
case the gauge symmetry may become reducible, the effect being most obvious for antisymmetric $p$-form fields. Moreover, for
reducible gauge symmetries there can be more than one gauge parameter at some level in general.

To be precise, a massless spin-$\Ss$ field $\Irrep{0}{\Ss}$ is defined by the exact sequence \be
\label{MSBGGFlat}\ComplexF{\Xi_p}{...}{\Xi_2}{\Xi_1}{\Verma{0}{\Ss}}{\Irrep{0}{\Ss}},\ee where $\Xi_r$ represents the gauge
symmetry at the level-$r$ \be\label{WELevelrSymmetry}\Xi_r=\bigoplus_{\substack{k_1+...+k_N=r\\
k_1=0,1;...;k_p=0,1\\s_1-k_1\geq s_2-k_2,...,s_{p-1}-k_{p-1}\geq s_p-k_p}}{\Verma{0}{\Y{s_1-k_1,...,s_p-k_p}}}.\ee

The number of gauge parameters at the first level is equal to the number of ways in which one cell can be removed from $\Ss$
without violating the Young conditions, i.e., it is equal to the number of groups of rows having equal length. There is only one
gauge parameter at the deepest level $r=p$ corresponding to $k_1=...=k_p=1$, whose Young diagram is obtained by removing one cell
from each row of $\Ss$, i.e. $\Y{s_1-1,s_2-1,...,s_p-1}$.

Due to the presence of gauge symmetry the physical degrees of freedom are no longer classified according to the representations
of $\msv$. The structure of invariant submodules is such that an irreducible representation of $\mls$ with the same symmetry
$\Ss$ is realized as an exact sequence of certain $\msv$-modules\footnote{In the simplest nontrivial case of a spin-one massless
field, by virtue of (\ref{MSBGGFlat}) an $\mls$ physical polarization vector $A_{\aI}$, $\aI=1,...,d-2$ is realized as an $\msv$
vector presented by the Maxwell potential $A_\mu$ subjected to $\square A_\mu=0$, so that $A_\mu$ reduces to a function of
$(d-1)$ variables, and $\pl^\mu A_\mu=0$, so that only $(d-1)$ of the $d$ components of $A_\mu$ are independent, modulo an $\msv$
scalar $\xi$, $\square\xi=0$ representing on-shell gauge symmetry $\delta A_\mu=\pl_\mu\xi$.}. Consequently, for $m^2=0$ the spin
degrees of freedom are in one-to-one correspondence with finite-dimensional irreducible representations of $\mls$.

From the group-theoretical point of view the construction of $\Irrep{m^2}{\Ss}$ is based on the well-know method of induced
representations, see classical work \cite{Wigner:1939cj} by Wigner for $d=4$ and \cite{Bekaert:2006py} for the review and
extension to arbitrary $d\geq4$.

\paragraph{Anti-de Sitter space, $\boldsymbol{\mathfrak{g}=\ads}$, \cite{Breitenlohner:1982bm, Nicolai:1984hb, Metsaev:1995re, Metsaev:1997nj, Brink:2000ag, Alkalaev:2009}.}
In the anti-de Sitter space the positive and negative frequency solutions of (\ref{FullSystemA}) can be separated. Therefore,
irreducible representations that are referred to as massive or (partially)-massless fields are identified with the
positive-frequency solutions of $\Verma{m^2}{\Ss}$ modulo certain pure gauge solutions in the (partially)-massless case.

For the (anti)-de Sitter case the appearance of gauge symmetry occurs at certain nonzero values of the mass parameter $m^2$,
which are measured in the units of the cosmological constant and hence tend to zero at the Minkowski limit. These critical values
of $m^2$ together with the structure of the gauge symmetries will be of main importance in what follows.

From the group-theoretical point of view the positive-frequency solutions of $\Verma{m^2}{\Ss}$ can be realized as a
Harish-Chandra module. The anti-de Sitter algebra $\mathfrak{g}=\ads$ admits a three-graded decomposition
$\mathfrak{g}=\mathfrak{g_{-1}}\oplus\mathfrak{g_{0}}\oplus\mathfrak{g_{+1}}$, i.e.,
$[\mathfrak{g_{0}},\mathfrak{g_{\pm1}}]\subset\mathfrak{g_{\pm1}}$ and
$[\mathfrak{g_{-1}},\mathfrak{g_{+1}}]\subset\mathfrak{g_{0}}$, with respect to its maximal compact subalgebra
$\mathfrak{g_{0}}=\AlgebraFont{so}(2)\oplus\msv$ of \ads. $\mathfrak{g_{-1}}$ and $\mathfrak{g_{+1}}$ are spanned by the
noncompact generators of \ads.

In order to construct a (unitary) irreducible representation of \ads\ one \cite{Nicolai:1984hb, deWit:1999ui} takes the vacuum
vector $|E_0,\Ss\rangle$ to be an irreducible representation of $\mathfrak{g_{0}}$, $E_0$ being the weight of
$\AlgebraFont{so}(2)$ and $\Ss$ being a Young diagram that characterizes an irreducible representation of \msv. The vacuum is
annihilated by $\mathfrak{g_{-1}}$, i.e. $\mathfrak{g_{-1}}|E_0,\Ss\rangle=0$. The module $\Verma{E_0}{\Ss}$ is freely generated
from $|E_0,\Ss\rangle$ by the positive grade generators $\mathfrak{g_{+1}}$, generic vector being
$\mathfrak{g_{+1}}\mathfrak{g_{+1}}...\mathfrak{g_{+1}}|E_0,\Ss\rangle$. $\Verma{E_0}{\Ss}$ is identified with the
positive-frequency solutions of $\Verma{m^2}{\Ss}$, where the lowest weights $E_0$, $\Ss$ of $\mathfrak{g}_0$ and the mass $m^2$
are related by \cite{Metsaev:1995re} \be m^2=\lambda^2\left(E_0(E_0-d+1)-s_1-...-s_p\right).\label{WEMassFormula}\ee

Given the mass $m^2$ and the spin $\Ss$ of a field, there are two roots $E_0^+$, $E_0^-$ of (\ref{WEMassFormula}) related by
$E_0^++E_0^-=d-1$, the maximal one $E_0^+$ corresponding to a massive or a (partially)-massless field and the minimal one
corresponding to its shadow partner \cite{Metsaev:2008fs}. The maximal root is meant hereinafter when referring to
(\ref{WEMassFormula}).

For certain values of the lowest energy $E_0$ there appears a singular vector, i.e. certain element $v$ of $\Verma{E_0}{\Ss}$
satisfies itself the condition of being vacuum $\mathfrak{g_{-1}} v=0$. Therefore, there appears a submodule
$\Verma{E_1}{\Ss_1}\subset\Verma{E_0}{\Ss}$ generated from $v$  $\mathfrak{g_{+1}}\mathfrak{g_{+1}}...\mathfrak{g_{+1}}v$, with
$E_1$ and $\Ss_1$ denoting the energy and the spin of $v$. From the field-theoretical point of view the equations
(\ref{FullSystemA})-(\ref{FullSystemE}) become invariant under certain gauge transformations with the gauge parameter having the
symmetry of $\Ss_1$. $E_0$ depends nontrivially on $\Ss_1$, hence, it is not possible to have two or more invariant submodules
$\Verma{E_1}{\Ss_1}$, $\Verma{E_2}{\Ss_2}$, ... simultaneously for the same value of $E_0$. Therefore \cite{Brink:2000ag,
Metsaev:1995re}, equations (\ref{FullSystemA})-(\ref{FullSystemE}) may have one gauge symmetry only as contrast to the Minkowski
case $\lambda^2=0$, in which all submodules appear at the same value of the mass parameter, $m^2=0$, and, hence, a generic
mixed-symmetry field has more than one gauge symmetry in Minkowski space. The precise determination of the possible gauge
symmetries is given in Section \ref{SGaugeFieldsdSAdS}.

There is no discrepancy between the number of degrees of freedom for massive fields in Minkowski and (anti)-de Sitter spaces
since the spin degrees of freedom of massive fields are classified according to representations of the same little algebra
$\msv$. As for massless fields, only those having all $s_i$ equal $s_1=s_2=...=s_p$, i.e., $\Ss$ is a rectangular diagram,
possess the same number of degrees of freedom both in Minkowski and (anti)-de Sitter spaces, these are the only fields for which
the number of gauge symmetries in Minkowski and (anti)-de Sitter spaces is equal. For instance, this is the case for symmetric
fields $p=1$ and antisymmetric fields $s_1=s_2=...=s_p=1$.

Partially-massless fields are nonunitary in the anti-de Sitter case and split in the Minkowski limit into a collection of
massless fields \cite{Zinoviev:2001dt, Alkalaev:2009}.

\paragraph{de Sitter space, $\boldsymbol{\mathfrak{g}=\ds}$.} The representation theory of the de Sitter algebra differs drastically
from that of the anti-de Sitter one. The de Sitter algebra mixes all solutions of (\ref{FullSystemA})-(\ref{FullSystemE}) into
one \ds-module, it not being possible to divide solutions of (\ref{FullSystemA})-(\ref{FullSystemE}) into positive and negative
frequency parts. Nevertheless, the notion of the lowest energy can be introduced \cite{Higuchi:1986wu}.

Despite these difficulties, gauge symmetry for (\ref{FullSystemA})-(\ref{FullSystemE}) appears at the same values of the mass as
determined for the anti-de Sitter case provided the change $\lambda^2\longrightarrow-\lambda^2$.

\section{Gauge Fields in \DSADS}\label{SGaugeFieldsdSAdS}
In this section we consider the general case of a spin-$\Ss$ field in (anti)-de Sitter space, where $\Ss$ is a finite-dimensional
irreducible bosonic representation of the (anti)-de Sitter 'Wigner little algebra' \msv, specified by a Young diagram
$\Ss=\Y{s_1,...,s_p}$, $p\leq[(d-1)/2]$. For $d=2n+1$ and $p=n$ (anti)-selfduality conditions have to be imposed to make the
representation irreducible. We prefer not to go into details concerning (anti)-selfdual representations and will ignore them.
Reviewed below are the results of \cite{Alkalaev:2009}, where gauge invariant wave equations in (anti)-de Sitter space are
studied, which generalizes numerous results of \cite{Higuchi:1986wu, Nicolai:1984hb, Metsaev:1995re, Metsaev:1997nj,
Brink:2000ag}.

\paragraph{Field theory, on-shell.} The field-theoretical statement is that given an irreducible field potential $\phi^\Ss\equiv\phi^{a(s_1),b(s_2),...,u(s_p)}$
having the symmetry of $\Ss=\Y{s_1,...,s_p}$, for any $q\in[1,p]$ provided $s_q-s_{q+1}>0$ and\footnote{It is convenient to set
$s_{p+1}=0$. The condition means that $q$ refers to a row from which at least one cell can be removed so that the resulted
picture is still a Young diagram.} any $t\in[1,s_q-s_{q+1}]$ there exists $m^2$ \be
m^2=\lambda^2\left((s_q-t-q)(d+s_q-t-q-1)-s_1-s_2-...-s_p\right)\label{MassFormula}\ee such that the wave equation
(\ref{FullSystemA}) for field $\phi^{\Ss}$ is invariant under the gauge transformations \be\delta
\phi^{a(s_1),b(s_2),...,u(s_p)}=\overbrace{D^c...D^c}^{t}\xi^{a(s_1),...,b(s_{q-1}),c(s_q-t),...,u(s_p)}+...\label{GCGaugeTransformations}\ee
where '...' stands for certain lower derivative terms and for the terms that restore the Young symmetry properties, if needed.
The gauge parameter is an irreducible tensor having the symmetry of $\Y{s_1,...,s_{q-1},s_q-t,s_{q+1},...,s_p}$. Gauge
transformations (\ref{GCGaugeTransformations}) are consistent with the transversality constraints (\ref{FullSystemB}), Young
symmetry conditions (\ref{FullSystemC}) and with the trace constraints (\ref{FullSystemD}-\ref{FullSystemE}) provided that the
gauge parameter itself is transverse, traceless and obeys the wave equation with \be
{m_\xi}^2=\lambda^2\left((s_q-q)(d+s_q-q-1)-s_1-s_2-...-s_p+t\right).\ee The order of derivative of gauge transformations is
equal to $t$, with $t=1$ and $t>1$ corresponding to massless fields and partially-massless fields, respectively. Important is
that no further extension of the gauge symmetry is possible. For $t=1$ the parameter $q$ refers to the Minkowski gauge symmetry
among $\Xi_1$ (\ref{WELevelrSymmetry}) that is allowed to survive in (anti)-de Sitter space.

Roughly speaking, for a given spin $\Ss$ there are as many different gauge fields as the ways in which a number of cells can be
removed from anyone row of $\Ss$ provided the resulted diagram is still a Young diagram (the length of a row is a nonincreasing
function of row).

\paragraph{Group theory.} Providing us with the description of the higher-level gauge symmetries, the group-theoretical statement is that given
an $\msv$-Young diagram $\Ss=\Y{s_1,...,s_p}$, for any $q\in[1,p]$ provided $s_q-s_{q+1}>0$ and any $t\in[1,s_q-s_{q+1}]$ there
exists the vacuum energy $E_0$ \be E_0(q,t)=d+s_q-t-q-1\label{EnergyFormula}\ee such that $\Verma{E_0}{\Ss}$ is reducible and the
irreducible representation $\Irrep{E_0}{\Ss}$, which is referred to as a massless or partially-massless field for $t=1$ and
$t>1$, respectively, is defined by the following exact sequence
\be0\rightarrow\Verma{E_q}{\Ss_q}\longrightarrow...\longrightarrow\Verma{E_1}{\Ss_1}\longrightarrow\Verma{E_0}{\Ss_0}\longrightarrow\Irrep{E_0}{\Ss_0}\rightarrow0\label{ExactSequence},\ee
where the lowest weights of $\AlgebraFont{so}(2)\oplus\msv$ are defined as \begin{align}E_i&=\begin{cases} d+s_q-t-q-1,& i=0,\\
d+s_{q-i+1}-(q-i+1)-1, & i=1,...,q,\end{cases}\label{AllLevelsE}\\
\Ss_i&=\begin{cases} \Y{s_1,...,s_p}\equiv\Ss,& i=0,\\
\Y{s_1,s_2,...,s_{q-1},s_q-t,s_{q+1},...,s_p}, & i=1,\\
\Y{s_1,...,s_{q-i},s_{q-i+2}-1,...,s_q-1,s_q-t,s_{q+1},...,s_p}, & i=2,...,q-1, \\
\Y{s_2-1,s_3-1,...,s_q-1,s_q-t,s_{q+1},...,s_p}, & i=q.\label{AllLevelsS}\end{cases}\end{align}

As contrast to the Minkowski case (\ref{WELevelrSymmetry}), there is only one gauge parameter/submodule at each level. The lowest
energy (\ref{EnergyFormula}) is related to the mass (\ref{MassFormula}) in accordance with (\ref{WEMassFormula}), and the same is
true for the gauge parameters/submodules of the exact sequence (\ref{ExactSequence}). If the field potential is taken to have the
symmetry of $\Ss_0\equiv\Ss$, as is implied throughout this paper, the gauge parameter at the level-$i$ has the symmetry of
$\Ss_i$.

The Casimir of $\Verma{E_0}{\Ss_0}$ and, if $E_0$ is one of the critical values (\ref{EnergyFormula}), of $\Irrep{E_0}{\Ss_0}$ is
given by \be C_2=E_0(E_0-d+1)+\sum_{i=1}^{i=p}s_i(d+s_i-2i-1).\ee

\paragraph{Towards an off-shell theory.} In order for gauge symmetry to be realized off-shell the trace constraints
(\ref{FullSystemD}-\ref{FullSystemE}) have to be relaxed, giving rise to the problem of extension of the field content. Indeed,
the trace constraints are not consistent with the relaxation of transversality constraints (\ref{FullSystemB}) for gauge
parameters\footnote{In principle, one may work in terms of traceless potentials and differentially constrained parameters
\cite{Alvarez:2006uu, Skvortsov:2007kz}. One more way to keep potentials irreducible is to impose the projector onto the
traceless part in the gauge transformations. However, in the latter case there exist no gauge invariant off-shell equations even
for totally-symmetric spin-$s$ fields.}. If the gauge symmetry is reducible similar arguments lead to the relaxation of trace
constraints for gauge parameters at deeper levels. Only the gauge parameter at the deepest level can be an algebraically
irreducible Lorentz tensor. For the case of Minkowski massless fields, the extension (\ref{FronsdalDoubleTrace}) for a spin-$s$
field was found by Fronsdal in \cite{Fronsdal:1978rb}, the extension for mixed-symmetry fields was conjectured by Labastida in
\cite{Labastida:1987kw}, recently proved to be correct in \cite{Bekaert:2006ix}. It has a simple interpretation within the
unfolded and frame-like approaches \cite{Skvortsov:2008sh, Skvortsov:2008vs}.

Because massive fields are not gauge fields no extension of the field content is needed for an off-shell version. However, to
construct a Lagrangian the field content has to be extended with the supplementary fields \cite{Fierz:1939ix, Singh:1974qz}. As
for the fields in (anti)-de Sitter space, the extension for $(\Ss,q=1,t=1)$ fields may be obtained from the frame-like
description of \cite{Alkalaev:2003qv}. As a by-product, we extend this result to the case of arbitrary-spin (partially)-massless
fields in (anti)-de Sitter space.

\section{Background Geometry}\label{SBackgroundGeometry}\setcounter{equation}{0}
In this Section we recall the description of the background geometry in terms of vielbein and Lorentz spin-connection, which can
be recognized as the Yang-Mills connections of the space-time symmetry algebra. For the case of (anti)-de Sitter space, whose
symmetry algebra is simple, there are additional simplifications.

\paragraph{Background geometry, Lorentz-covariantly.} As is well-known, instead of working with the metric tensor $g_{\mu\nu}$
one \cite{Weyl:1929fm} may introduce a nonholonomic basis defined by a nonsingular matrix $h^a_\mu$, called tetrad/vielbein/frame
field. The index $a$ of the tetrad $h^a_\mu$ is a Lorentz one, it is raised and lowered with the invariant tensor $\eta_{ab}$ of
the Lorentz algebra. To define a covariant derivative the Lorentz spin-connection $\varpi^{a,b}_\mu=-\varpi^{b,a}_\mu$ is to be
introduced. Major achievement is in that $\varpi^{a,b}\equiv\varpi^{a,b}_\mu dx^\mu$ and $h^a\equiv h^a_\mu dx^\mu $ were
recognized \cite{Kibble:1961ba, MacDowell:1977jt, Stelle:1979aj} to be the Yang-Mills fields associated with the generators
$L_{a,b}$ and $P_a$ of Lorentz rotations and translations, respectively. Generators $L_{a,b}$ and $P_a$ form $\iso$, $\ds$ or
$\ads$. The Minkowski, de Sitter or anti-de Sitter background geometry can be described by the zero curvature (flatness)
condition $d\Omega+[\Omega,\wedge\Omega]=R^{a,b}L_{a,b}+T^aP_a=0$ for the Yang-Mills connection
$\Omega=\varpi^{a,b}L_{a,b}+h^aP_a$, where
\begin{align}
        &&&T^a&&=&&dh^a+\varpi^{a,}_{\phantom{a,}b}\wedge h^b=0,&&&&\label{UnfldIsoFlatnessA}\\
        &&&R^{a,b}&&=&&d\varpi^{a,b}+\varpi^{a,}_{\phantom{a,}c}\wedge\varpi^{c,b}\pm\lambda^2h^a\wedge h^b=0.&&&&\label{UnfldIsoFlatnessB}
\end{align}
On condition that $h^a_\mu$ is a nonsingular matrix, any solution of (\ref{UnfldIsoFlatnessA}-\ref{UnfldIsoFlatnessB}) describes
Minkowski ($\lambda^2=0$), de Sitter ($+\lambda^2$) or anti-de Sitter\footnote{In the expressions similar to
(\ref{UnfldIsoFlatnessB}), upper/lower sign corresponds to the de Sitter/anti-de Sitter case hereinafter.} ($-\lambda^2$)
geometry and provides us with the basis of a fiber space $h^a_\mu$ and with Lorentz spin-connection $\varpi^{a,b}_\mu$.

For the case of the Minkowski geometry a simple solution of (\ref{UnfldIsoFlatnessA}-\ref{UnfldIsoFlatnessB}) is given by
Cartesian coordinates $h^a_\mu=\delta^a_\mu$, $\varpi^{a,b}_\mu=0$. It is assumed further that $h^a_\mu$ and $\varpi^{a,b}_\mu$
obey (\ref{UnfldIsoFlatnessA}-\ref{UnfldIsoFlatnessB}) but the advantage is that no explicit solution is needed either to write
down field-equations or to construct actions, which is the most effective for the (anti)-de Sitter case \cite{Lopatin:1987hz,
Vasiliev:2001wa, Alkalaev:2003qv, Alkalaev:2005kw, Alkalaev:2006rw}.

With the help of $\varpi^{a,b}_\mu$ one defines the Lorentz covariant derivative of differential forms with values in any
finite-dimensional representation of \lorentz, i.e., having some fiber Lorentz indices, e.g., for a degree-$q$ form\footnote{In
what follows the form degree is indicated by the bold subscript, except for the connections describing the background geometry,
and the wedge symbol $\wedge$ is omitted.} $T^{ab...}\fm{q}\equiv
T^{ab...}_{\mu_1\mu_2...\mu_q}dx^{\mu_1}dx^{\mu_2}...dx^{\mu_q}$ \be \DL
T^{ab...}\fm{q}=dT^{ab...}\fm{q}+\varpi^{a,}_{\phantom{a,}m}T^{mb...}\fm{q}+\varpi^{b,}_{\phantom{b,}m}T^{am...}\fm{q}+... \,\,
.\ee

\paragraph{Background geometry, \DSADS-covariantly \cite{Stelle:1979aj}.} Since the (anti)-de Sitter algebra is simple and there
exists an invariant tensor $\eta_{AB}$, (\ref{UnfldIsoFlatnessA}-\ref{UnfldIsoFlatnessB}) is simplified to \be
d\Omega^{A,B}+\Omega^{A,}_{\phantom{A,}C}\Omega^{C,B}=0\label{AntideSitterFlatness},\ee where $\Omega^{A,B}_\mu
dx^\mu=-\Omega^{B,A}_\mu dx^\mu$, $A,B,...=0,1,...,d$, is a connection of the (anti)-de Sitter algebra. The Lorentz covariant
equations (\ref{UnfldIsoFlatnessA}-\ref{UnfldIsoFlatnessB}) can be recovered from (\ref{AntideSitterFlatness}) if one makes
identifications
\begin{align} \Omega^{a,}_{\phantom{a,}\bullet}=\lambda h^a, && \Omega^{a,b}=\varpi^{a,b},\label{UnfldSplitting}\end{align}
where $\bullet$ denotes the extra value of the \ads{} or \ds{} vector index as compared to the \lorentz{} vector index, i.e.
$A=a,\bullet$; $a=0,1,...,d-1$; $\bullet=d$.

The splitting (\ref{UnfldSplitting}) can be made manifestly \dSAdS-covariant \cite{Stelle:1979aj, Vasiliev:2001wa} if one
identifies the Lorentz algebra as a stability algebra of a vector compensator field $V^A$, which is convenient to normalize to
unit length, \be V^BV_B=\mp 1.\label{UnfldCompensatorNorm}\ee The generalized vielbein field $E^A_\mu dx^\mu$\begin{align}
\lambda E^A=\DO V^A=dV^A+\Omega^{A,}_{\phantom{A,}B}V^B\label{UnfldGeneralizedE}
\end{align}
is assumed to have the maximal rank, which is $d$. Therefore, $E^A$ defines a nonsingular vielbein field orthogonal to $V^A$
inasmuch as $E^BV_B=0$ by virtue of (\ref{UnfldGeneralizedE}) and (\ref{UnfldCompensatorNorm}). The Lorentz covariant derivative
$\DL=d+\Omega_L$ is defined with respect to the Lorentz connection $\Omega_L^{A,B}$ \be\Omega_L^{A,B}=\Omega^{A,B}\mp
\lambda(V^AE^B-E^AV^B).\label{BGLorentzConnection}\ee  Both the compensator and the generalized vielbein are Lorentz-covariantly
constant \be\DL V^A=0, \qquad \DL E^A=0.\label{BGCovariantConstancy}\ee

One can always choose the 'standard gauge' for the compensator field $V_A=\delta^A_{\bullet}$, then $\lambda
E^A=\Omega^{A,}_{\phantom{A,}\bullet}$, $E^\bullet_\mu=0$ and $\Omega_L^{a,b}=\Omega^{a,b}$, which coincides with
(\ref{UnfldSplitting}).

It is worth stressing that the flatness condition (\ref{AntideSitterFlatness}) can simply be rewritten as $\DO^2=0$.

\section{Gauge connections of (anti)-de Sitter algebra}\label{SGaugeConnections}\setcounter{equation}{0}

Having (anti)-de Sitter space as a background, worth being scrutinized thoroughly are the generalized Yang-Mills connections of
the (anti)-de Sitter algebra that are differential forms of arbitrary degree with values in any finite-dimensional module of the
(anti)-de Sitter algebra.

Let $W^{AB...}\fm{q}$ be a $q$-form over $\dSAdS$ with values in any tensor representation of the (anti)-de Sitter algebra, i.e.
having some fiber tensor indices $A,B,...$ ranging $0,...,d$. Fiber indices may have some symmetry and/or trace properties
ensuring algebraic irreducibility, if needed. With the help of the flat background connection $\Omega^{A,B}$ one defines the
(anti)-de Sitter covariant derivative\footnote{Spin-tensors can be considered on equal footing, the covariant derivative contains
an extra term $\frac18\Omega^{A,B}[\gamma_A,\gamma_B]$, where $\gamma_A$ are the generators of the Clifford algebra
$\gamma_A\gamma_B+\gamma_B\gamma_A=2\eta_{AB}$. } $\DO$ of $W^{AB...}\fm{q}$ \be\DO W^{AB...}\fm{q}=
dW^{AB...}\fm{q}+\Omega^{A,}_{\phantom{A,}M}W^{MB...}\fm{q}+\Omega^{B,}_{\phantom{B,}M}W^{AM...}\fm{q}+...\quad.\ee $\DO$
preserves symmetry and/or trace properties.

Given a $q$-form $W^{AB...}\fm{q}$ one may introduce the $(q+1)$-form field strength $R^{AB...}\fm{q+1}=\DO W^{AB...}\fm{q}$,
which has the same symmetry/trace properties as $W^{AB...}\fm{q}$. The field strength turns out to be invariant under the gauge
transformations $\delta W^{AB...}\fm{q}=\DO \xi^{AB...}\fm{q-1}$ by virtue of the flatness condition $\DO^2=0$
(\ref{AntideSitterFlatness}), where the gauge parameter is a $(q-1)$-form with values in the same module as $W^{AB...}\fm{q}$. By
the same reason $W^{AB...}\fm{q}$ is invariant under the second level gauge transformations $\delta \xi^{AB...}\fm{q-1}=\DO
\xi^{AB...}\fm{q-2}$ and so on until $\delta \xi^{AB...}\fm{1}=\DO \xi^{AB...}\fm{0}$. In addition, the field strength satisfies
the Bianchi identities $\DO R^{AB...}\fm{q+1}=0$.

As we have already stated in the introduction, if the form degree and the symmetry/trace properties of the fiber indices are
chosen properly, $W^{AB...}\fm{q}$ is a natural framework for describing gauge fields in (anti)-de Sitter space, the idea
suggested first in \cite{Vasiliev:2001wa} for $(\Y{s},1,1)$ fields, in \cite{Alkalaev:2003qv} for the $(\Ss,1,1)$ fields and in
\cite{Skvortsov:2006at} for $(\Y{s},1,t)$ fields. It has a nice property of being manifestly (anti)-de Sitter covariant. A single
$q$-form connection incorporates the whole set of physical and auxiliary Lorentz fields, which is obtained by taking various
projections with respect to the compensator $V^C$.

Whereas it is sufficient to give consideration only to irreducible representations, in what follows all differential forms take
values in irreducible tensor representations of the (anti)-de Sitter algebra, leaving spin-tensor representations out of the key
target of the paper. This means that (i) the fiber indices have the symmetry of some Young diagram; (ii) the contraction of any
two fiber indices with the invariant tensor $\eta_{AB}$ of the (anti)-de Sitter algebra vanishes identically, i.e. fiber tensors
are traceless.

As we have already done for Lorentz tensors, it is convenient to take all tensors in the symmetric basis, meaning that tensor
indices consist of groups with the manifest symmetry among the indices from any group\footnote{All results obtained in the paper
do not depend on the choice of a basis for mixed-symmetry tensors, of course. Instead of the separation of indices into groups of
symmetric ones, one may single out the groups of anti-symmetric indices.}. For instance, a $q$-form with values in the
irreducible tensor representation $\Ds$ of the (anti)-de Sitter algebra that is characterized by Young diagram
$\Ds=\Y{s_1,...,s_n}$ \be\label{GCSample}
W^{A(s_1),B(s_2),...,U(s_n)}_{\mu_1\mu_2...\mu_q}dx^{\mu_1}dx^{\mu_2}...dx^{\mu_q}\equiv W^\Ds\fm{q}\ee is symmetric in each
group of indices $A_1...A_{s_1}$, ..., $U_1...U_{s_n}$, satisfies the Young symmetry condition\footnote{As before, a group of
symmetric indices is denoted by one letter, the number of symmetric indices placed in brackets; indices from different groups
denoted by the same letter are to be symmetrized.} \be W^{A(s_1),...,B(s_i),...,BC(s_j-1),...,U(s_n)}\fm{q}\equiv0,\qquad
i,j=1,...,n,\quad i<j\ee and the contraction of any two fiber indices with $\eta_{CD}$ is identically zero.

As illustrated below, any manifestly (anti)-de Sitter covariant formulation in terms of some gauge connection $W^\Ds\fm{q}$ can
be demoted first to the Lorentz-covariant frame-like formulation by decomposing the (anti)-de Sitter module $\Ds$ into
irreducible Lorentz modules with the help of the compensator $V^C$, with a collection of $q$-form connections of the Lorentz
algebra arising at this stage. One of those Lorentz connections is the generalized frame-like field that incorporates the
dynamical metric-like field $\phi^\Ss$. The rest of fields are various generalized Lorentz connections representing auxiliary
fields. Then, it can be demoted even further, to the metric-like formulation, by converting all differential forms with fiber
indices of the Lorentz algebra, obtained at the first stage, into fully world or fully fiber tensors with the help of the
background vielbein $h^a_\mu$ or its inverse $h^\mu_a$; and, then, by fixing the vast algebraic gauge symmetry that we will see
is present in the theory.
\be\stackTB{\dSAdS-covariant}{frame-like}\xrightarrow{V^A}\stackTB{Lorentz-covariant}{frame-like}\xrightarrow[h^a_\mu]{h^{\mu}_
a}\stackTB{Lorentz-covariant}{metric-like}\nonumber\ee For instance, in the case of a massless spin-$(s\geq2)$ field the demotion
sequence is shown in the introduction (\ref{IntroDemotionSequence}).

\paragraph{From \DSADS-covariant to Lorentz-covariant, dimensional reduction.} The Lorentz algebra is
defined as the subalgebra of the (anti)-de Sitter algebra $\mathfrak{g}$ annihilating the compensator $V^C$. The result of the
restriction of an irreducible $\mathfrak{g}$-module $\Ds=\Y{s_1,...,s_n}$ to its Lorentz subalgebra is easy to formulate in terms
of Young diagrams \be\mbox{Res}^{\mathfrak{g}}_{\lorentz}
\Ds\longrightarrow\bigoplus_{k_1=s_2}^{k_1=s_1}...\bigoplus_{k_{n-1}=s_{n}}^{k_{n-1}=s_{n-1}}\bigoplus_{k_n=0}^{k_n=s_n}
\Ds_{k_1,...,k_{n-1},k_n}\label{GCBranching},\ee where $\Ds_{k_1,...,k_{n-1},k_n}=\Y{k_1,...,k_{n-1},k_n}$. Thus, the result of
the restriction of $\Yy$ is given by various Young diagrams obtained by removing an arbitrary (possibly zero) number of cells
from the right of rows of $\Yy$ provided that each truncated row is not shorter than the next row of the initial diagram $\Yy$.
It is also useful to introduce a $V$-grade $g$ that is equal to $k_1+...+k_n-s_2-...-s_p$ for the element
$\Y{k_1,...,k_{n-1},k_n}$, so that $g=0$ for the element of the lowest rank and $g=s_1$ for the element of the highest rank. The
Lorentz subalgebra leaves each $\Y{k_1,...,k_n}$ invariant. The translation generators act between different $\Y{k_1,...,k_n}$,
mapping a grade-$g$ module to the modules at grade $(g\pm1)$.

Therefore, the $\dSAdS$ gauge connection $W^\Ds\fm{q}$ is reduced to a collection of gauge connections of the Lorentz algebra,
which is $V$-graded,
\begin{align} &W^\Ds\fm{q} &&\longleftrightarrow&& \omega^{\Ds_{k_1,...,k_n}}\fm{q},\quad
\substack{\displaystyle k_1=s_2,...,s_1,\\ \displaystyle\phantom{k_1}...,\\ \displaystyle k_{n-1}=s_n,...,s_{n-1},\\
\displaystyle k_n=0,...,s_n.}
\end{align} It is obvious that an irreducible tensor $R^{\Xx}$ of the $\dSAdS$-algebra that is fully orthogonal to the
one-dimensional subspace defined by the compensator is equivalent to an irreducible tensor of the Lorentz algebra that is defined
by the same Young diagram \Xx. Therefore, the irreducible $\dSAdS$-tensor $T^\Ds$ has the decomposition into irreducible tensors
of the Lorentz algebra of the form \be\label{GCTensorDecomposition}
T^{\Ds}=\sum_{k_1,...,k_p}\left(\overbrace{V...V}^{s_1-k_1}...\overbrace{V...V}^{s_n-k_n}T^{\Ds_{k_1,...,k_n}}+\mbox{perm}+\eta\right),\ee
where each tensor $T^{\Ds_{k_1,...,k_n}}$ of the (anti)-de Sitter algebra is fully orthogonal to $V^C$, i.e. the contraction of
any index with $V^C$ vanishes. '$\mbox{perm}$' stands for the terms with permuted indices and $\eta$ stands for the terms with
$\eta_{AB}$, which are present in general since $T^{\Ds}$ is subjected to certain symmetry and trace conditions\footnote{For
example, a rank-three traceless tensor $T^{AA,B}$ having the symmetry of $\smallboldpic{\YoungBA}$ decomposes as
$T^{AA,B}=R^{AA,B}+V^AR^{A,B}+\left[V^AR^{AB}-V^BR^{AA}\right]+
\left[\left(V^AV^AR^B-V^AV^BR^A\right)-\frac1dV^CV_C\left(\eta^{AA}R^B-\eta^{AB}R^A\right)\right]$, where $R^{AA,B}$, $R^{A,B}$,
$R^{AB}$ and $R^A$ are irreducible tensors orthogonal to $V^C$ having the symmetry of $\smallboldpic{\YoungBA}$,
$\smallboldpic{\YoungAA}$, $\smallboldpic{\YoungB}$ and $\smallboldpic{\YoungA}$, respectively. An equivalent statement is that
$T^{AA,B}$ decomposes into irreducible Lorentz tensors $R^{aa,b}$, $R^{a,b}$, $R^{aa}$ and $R^a$ that have the symmetry of
$\smallpic{\YoungBA}$, $\smallpic{\YoungAA}$, $\smallpic{\YoungB}$ and $\smallpic{\YoungA}$ and have grade $2$, $1$, $1$ and $0$,
respectively.}.

Let us consider some technical details that allow to perform the reduction to the Lorentz-covariant expressions explicitly in
terms of tensors.

In tensorial terms, to get the element $T^{\Ds_{k_1,...,k_n}}$ one contracts $(s_i-k_i)$ compensators with the $i$-th group of
the fiber indices of $T^\Ds$ \vspace{-0.4cm}\be\label{GCProjectionA}
T^{A(k_1)A'(s_1-k_1),B(k_2)B'(s_2-k_2),...,U(k_n)U'(s_n-k_n)}\overbrace{V_{A'}...V_{A'}}^{s_1-k_1}\overbrace{V_{B'}...V_{B'}}^{s_2-k_2}.....\overbrace{V_{U'}...V_{U'}}^{s_n-k_n}.\ee
To simplify notation any index contracted with the compensator will be denoted by $\bullet$, which is done on account of the fact
that we can always choose the standard gauge for $V^A$, as in (\ref{UnfldSplitting}). In the standard gauge, any Lorentz tensor
$r^{a(k_1),...,u(k_n)}$ can simply be embedded into the tensor $R^{A(k_1),...,U(k_n)}$ of the (anti)-de Sitter algebra,
$R^{a(k_1),...,u(k_n)}=r^{a(k_1),...,u(k_n)}$ and $R^{A(k_1),...,C(k_i-1)\bullet,...,U(k_n)}=0$ for any $i=1,...,n$.

Therefore, instead of working with $V$-orthogonal tensors of the (anti)-de Sitter algebra we can explicitly work in terms of
tensors of the Lorentz algebra, for example, in the standard gauge to get $T^{\Ds_{k_1,...,k_n}}$ one
writes\be\label{GCRawTensor} T^{a(k_1)\bullet(s_1-k_1),b(k_2)\bullet(s_2-k_2),...,u(k_n)\bullet(s_n-k_n)}.\ee Despite having the
correct number of fiber indices in each group, (\ref{GCProjectionA}) and (\ref{GCRawTensor}) generally neither have definite
Young symmetry nor are orthogonal to $V^C$. On account of this, let us refer to (\ref{GCRawTensor})-like expressions as 'raw'
ones. In order to single out the irreducible Lorentz tensor having the symmetry of $\Ds_{k_1,...,k_n}$, (\ref{GCProjectionA}) and
(\ref{GCRawTensor}) have to be supplemented with certain '$\mbox{perm}$'- and $\eta$-terms.

It is worth noting that any 'raw' fiber tensor of the form (\ref{GCRawTensor}) is not generally traceless with respect to the
Lorentz invariant tensor $\eta_{ab}$. Any contraction of two Lorentz indices in (\ref{GCRawTensor}) is equivalent to the
contraction of two more compensators modulo the sign factor, which is $(-)+$ for (anti)-de Sitter space.

Note also that the contraction of more than $(s_i-s_{i+1})$ compensators with the $i$-th group of indices may not vanish
identically, it can be expressed as certain sum of the terms having no more than $(s_i-s_{i+1})$ compensators contracted with the
$i$-th group.

There are cases for which no '$\mbox{perm}$'-terms are needed, so that contracted with the compensators 'raw' tensor itself
satisfies Young conditions. As to (\ref{GCSample})
\begin{Lemma}[A] Provided that the $i$-th group of indices, $i=k,...,n$ is contracted with $s_{i}-s_{i+1}$ ($s_n$ for $i=n$;
$s_{i}-s_{i+1}$ may be zero) compensators the resulting tensor has the symmetry of
$\Y{s_1,...,s_{k-1},s_{k+1},s_{k+2},...,s_{n}}$, i.e., as if the $k$-th row is dropped off, and it is Lorentz-traceless with
respect to the indices of the groups $k,...,n-1$. \be
r^{a(s_1),..,b(s_{k-1}),c(s_{k+1}),...,u(s_{n-1})}=T^{a(s_1),...,b(s_{k-1}),c(s_{k+1})\bullet(s_k-s_{k+1}),...,u(s_{n})\bullet(s_{n-1}-s_n),\bullet(s_n)}\nonumber.\ee
\end{Lemma}

Whereas all manifestly $\dSAdS$-covariant expressions, e.g., the gauge transformation law, involve the covariant derivative $\DO$
only, to reinterpret any (anti)-de Sitter covariant expression in terms of the Lorentz subalgebra it is convenient to extract the
Lorentz-covariant derivative $\DL$ out of $\DO$ according to (\ref{BGLorentzConnection})
\begin{align}&\delta W^{AB...}\fm{q}&&=&&\DL \xi^{AB...}\fm{q-1}&&\pm&&\lambda V^A E_M \xi^{MB...}\fm{q-1}&&\mp&&\lambda E^A V_M
\xi^{MB...}\fm{q-1}&&+&&...\quad,\label{GCDOSlitting}\end{align} and in a similar manner for any other expressions.

By virtue of (\ref{BGCovariantConstancy}), the decomposition (\ref{GCTensorDecomposition}) and the property of being orthogonal
to $V^C$ are preserved by the action of $\DL$ rather than $\DO$. Besides $\DL$ there are two more operators on the {\it r.h.s.}
of (\ref{GCDOSlitting}). The first one $ V^{..} E_M$ with a free index on the compensator decreases the grade by one, and the
second one $E^{..} V_M$, which contracts the compensator with the index of the field, increases the grade by one.

In terms of 'raw' fields (\ref{GCRawTensor}) and the signs for the anti-de Sitter case, (\ref{GCDOSlitting}) reads
\begin{align}
\delta &W^{a(k_1)\bullet(s_1-k_1),...,u(k_n)\bullet(s_n-k_n)}\fm{q}=\DL\xi^{a(k_1)\bullet(s_1-k_1),...,u(k_n)\bullet(s_n-k_n)}\fm{q-1}+\nonumber\\
&\qquad\qquad+\lambda\sum_{i=1}^{i=n}(s_i-k_i)h_m\xi^{a(k_1)\bullet(s_1-k_1),...,c(k_i)m\bullet(s_i-k_i-1),...,u(k_n)\bullet(s_n-k_n)}\fm{q-1}+\nonumber\\
&\qquad\qquad-\lambda\sum_{i=1}^{i=n}h^c\xi^{a(k_1)\bullet(s_1-k_1),...,c(k_i-1)\bullet(s_i-k_i+1),...,u(k_n)\bullet(s_n-k_n)}\fm{q-1},\label{GCLorentzGaugeLaw}
\end{align}
where prefactor $(s_i-k_i)$ is due to the identical permutations of the indices contracted with the compensator. Instead of 'raw'
fields one can single out irreducible fields \be
\omega^{a(k_1),...,u(k_n)}\fm{q}=\boldsymbol{\Pi}\left(W^{a(k_1)\bullet(s_1-k_1),...,u(k_n)\bullet(s_n-k_n)}\fm{q}\right),\ee
where $\Pi$ is a projector containing '$\mbox{perm}$'- and $\eta$-terms such that all traces and symmetry components other than
$\Y{k_1,...,k_n}$ are removed. Rewritten in terms of irreducible Lorentz fields, (\ref{GCLorentzGaugeLaw}) reads
\begin{align}
\delta
\omega^{a(k_1),...,u(k_n)}\fm{q}&=\DL\xi^{a(k_1),...,u(k_n)}\fm{q-1}+\label{GCLorentzGaugeLawIrr}\\+&\lambda\boldsymbol{\Pi}\left(\sum_{i=1}^{i=n}h_m
\xi^{a(k_1),...,c(k_i)m,...,u(k_n)}\fm{q-1}\right)-\lambda\boldsymbol{\Pi}\left(\sum_{i=1}^{i=n}h^c\xi^{a(k_1),...,c(k_i-1),...,u(k_n)}\fm{q-1}\right)\nonumber,
\end{align}
where we omit certain nontrivial coefficients in front of $\boldsymbol{\Pi}$. The first and the second operators in the second
line take a field with the symmetry of $\Y{k_1,...,k_i\pm1,...,k_n}$ to the field with the symmetry of $\Y{k_1,...,k_i,...,k_n}$,
these operators are called $\Sigm$ and $\Sigp$, respectively. $\boldsymbol{\sigma_-}$ and $\boldsymbol{\sigma_+}$ are the
operators $V^{..}E_M$ and $E^{..}V_M$ from (\ref{GCDOSlitting}) in terms of the irreducible Lorentz components.

As it can be seen either from (\ref{GCLorentzGaugeLaw}) or from (\ref{GCLorentzGaugeLawIrr}) the gauge symmetry has both the
differential and the algebraic(Stueckelberg) parts. The latter can be used to gauge away certain components of the Lorentz
connections $\omega^{...}\fm{q}$. By the same reason not all of the gauge parameters $\xi^{...}\fm{q-1}$ do affect
$\omega^{...}\fm{q}$ because of the reducibility of gauge symmetry.

\paragraph{From Lorentz frame-like to metric-like.}

Suppose we are given a degree-$q$ form with values in some irreducible tensor representation $\Xx=\Y{k_1,...,k_n}$ of the Lorentz
algebra \be\omega^{a(k_1),b(k_2),...,u(k_n)}_{\mu_1...\mu_q}dx^{\mu_1}....dx^{\mu_q}.\label{GCLorentzConnection}\ee With the help
of the inverse background vielbein $h^{a\mu}$, $h^{a\mu}h^b_{\mu}=\eta^{ab}$, all world indices can be converted to fiber ones
(or vice-versa with the help of $h_{a\mu}$)
\be\omega^{a(k_1),b(k_2),...,u(k_n)|v_1...v_q}=\omega^{a(k_1),b(k_2),...,u(k_n)}_{\mu_1...\mu_q}h^{v_1\mu_1}...h^{v_q\mu_q}.\label{GCLorentzConnectionFiber}\ee
The fully fiber tensor is obviously antisymmetric in indices $v_1,...,v_q$. Since there are no algebraic conditions between
indices $a(k_1),...,u(k_n)$ and indices $v_1,..,v_q$, to interpret $\omega^{a(k_1),b(k_2),...,u(k_n)|v_1...v_q}$ in terms of
irreducible Lorentz tensors is equivalent to taking the $\lorentz$-tensor product\vspace{-0.3cm} \be\Xx\otimes_\lorentzS
\Y{\overbrace{1,...,1}^q}\label{FormsTensorProduct}\ee of $\Xx$ with a one column diagram of height $q$, which represents
antisymmetric indices $v_1...v_q$.

The simplest way to obtain a degree-$q$ form with fiber indices having the symmetry of $\Xx$ is to take a degree-zero form
$C^{\Zz}$ with fiber indices having the symmetry of $\Zz=\Y{k_1+1,...,k_q+1,k_{q+1},...,k_n}$
\be\omega^{a(k_1),b(k_2),...,u(k_n)}_{\mu_1...\mu_q}=C^{a(k_1)v_1,b(k_2)v_2,...,c(k_q)v_q,...,u(k_n)}h_{v_1\mu_1}....h_{v_q\mu_q},\label{FormsSimplest}\ee
which is equivalent to the statement that (\ref{FormsTensorProduct}) contains $\Zz$. Due to the anticommutativity of $h^a$,
(\ref{FormsSimplest}) has automatically the symmetry of $\Xx$, i.e. no Young symmetrizers are needed in the symmetric basis.

In spite of the fact that the $\dSAdS$ connection $W^\Ds\fm{q}$ gives rise to a large number of Lorentz connections, which in its
turn give rise to an even larger number of metric-like Lorentz tensors, all physically relevant components are obtained by virtue
of
\begin{Lemma}[B] Given (\ref{GCLorentzConnection}) and its fiber version (\ref{GCLorentzConnectionFiber}), the fiber tensor \be
B^{a_1(k_1+1),...,a_q(k_q+1),b(k_{q+1}),...,u(k_n)}=\omega^{a_1(k_1),...,a_q(k_q),b(k_{q+1}),...,u(k_n)|a_1...a_q}\ee has the
symmetry of $\Zz$. Despite having definite Young symmetry $B^{...}$ is not completely traceless, instead the trace properties are
\begin{align}
&\eta_{cc}\eta_{cc}B^{a_1(k_1+1),...,cccc a_i(k_i-4),...,a_q(k_q+1),b(k_{q+1}),...,u(k_n)}\equiv0, && i=1...q\label{GCDoubleTracelessnss}\\
&\eta_{cc}B^{a_1(k_1+1),...,a_q(k_q+1),b(k_{q+1}),...,ccf(k_j-2),...,u(k_n)}\equiv0, && j=q+1...n\label{GCTracelessnss}
\end{align}
\end{Lemma}
Consequently, $B^{...}$ satisfies the Fronsdal-Labastida double-trace constraints for the first $q$ groups of indices, and is
traceless in the rest of the indices. Therefore, the Labastida-like constraints seem to have come from certain Lorentz
connections \cite{Skvortsov:2008vs}.

The irreducible component of $B^{...}$ with the highest rank, i.e. \Zz, (the highest weight part of (\ref{FormsTensorProduct}))
will be of main interest for us because it will be identified with the physical field $\phi^{\Ss_0}$, and with the gauge
parameters thereof $\xi^{\Ss_1}$, ..., $\xi^{\Ss_q}$.

\section{Gauge fields vs. Gauge connections}\label{SGeneralCase}\setcounter{equation}{0}
In order to describe a spin-$\Ss$, $\Ss=\Y{s_1,...,s_p}$, (partially)-massless field $\phi^{\Ss}$ whose gauge parameter
$\xi^{\Ss_1}$ has the symmetry of $\Ss_1=\Y{s_1,...,s_q-t,s_{q+1},...,s_p}$, i.e. it is obtained by removing $t$ boxes from the
$q$-th row of $\Ss$, let us consider a $q$-form $W^\Ds\fm{q}$ with values in the irreducible tensor representation $\Ds$ of the
(anti)-de Sitter algebra \be\Ds\equiv\Ds(\Ss,q,t)=\Y{s_1-1,...,s_q-1,s_q-t,s_{q+1},...,s_p}\label{GFGCAdSDiagram}.\ee So that in
order to build the Young diagram $\Ds$ of the (anti)-de Sitter algebra from the Young diagram $\Ss$ of the Wigner little algebra
one removes one cell from the right of rows $1,2,...,q$, inserts after the $q$-th row an extra row of length $(s_q-t)$, the rest
of the rows of $\Ss$ remains untouched. In symmetric basis the gauge field reads explicitly as \be
W^{A_1(s_1-1),...,A_q(s_q-1),B(s_q-t),C(s_{q+1}),...,U(s_{p})}_{\mu_1\mu_2...\mu_q} dx^{\mu_1}\wedge dx^{\mu_2}\wedge...\wedge
dx^{\mu_q}.\ee

The gauge transformations at all levels of reducibility together with the manifestly gauge invariant field strength, satisfying
certain Bianchi identities, can be written immediately with the help of a flat connection $\DO$
\begin{align}
\DO R^{\Ds}\fm{q+1}&=0,\nonumber\\
R^{\Ds}\fm{q+1}&=\DO W^{\Ds}\fm{q}, & \delta R^{\Ds}\fm{q+1}&=0,\nonumber\\
\delta W^{\Ds}\fm{q}&=\DO \xi^{\Ds}\fm{q-1},&&\label{MainGaugeTransformations}\\
\delta \xi^{\Ds}\fm{q-1}&=\DO \xi^{\Ds}\fm{q-2},&&\nonumber\\
...&=...,&&\nonumber\\
\delta \xi^{\Ds}\fm{1}&=\DO \xi^{\Ds}\fm{0}.&&\nonumber
\end{align}

We will demonstrate that there exists the following 'embedding' $\Verma{E_0}{\Ss_0}\rightarrow\phi^{\Ss_0}\rightarrow
e^{\Ll_0}\fm{q}\rightarrow W^{\Ds}\fm{q}$, i.e., $W^{\Ds}\fm{q}$ decomposes into a collection of the connections of the Lorentz
algebra, among which is $e^{\Ll_0}\fm{q}$, referred to as the generalized frame, that contains the metric-like dynamical field
$\phi^{\Ss_0}$ with the symmetry of $\Ss_0\equiv\Ss$ and provided that certain components of the field strength are set to zero
$\phi^{\Ss_0}$ satisfies the wave equation with the correct mass-like term (\ref{MassFormula}), which is determined by $E_0$,
$\Ss$ (\ref{EnergyFormula}). Analogously for the level-$i$ gauge parameter $\xi^{\Ss_i}$,
$\Verma{E_i}{\Ss_i}\rightarrow\xi^{\Ss_i}\rightarrow \xi^{\Ll_i}\fm{q-i}\rightarrow \xi^{\Ds}\fm{q-i}$.

\paragraph{\DSADS\ down to Lorentz Frame-like.}
It is useful to introduce $\gamma_i$, $i=1,...,p+1$
$$\gamma_i=\begin{cases} s_i-s_{i+1}, & i=1,...,q-1,\\
t-1, & i=q,\\
s_q-s_{q+1}-t, & i=q+1,\\
s_{i-1}-s_i, & i=q+2,...,p,\\
s_p, &i=p+1,\end{cases}$$ that is defined as the difference between the length of the $i$-th and the $(i+1)$-th row of $\Ds$,
i.e., it is equal to the maximal number of the compensators that can be contracted with the $i$-th group of indices of
$W^\Ds\fm{q}$ according to the restriction rule (\ref{GCBranching}), and it is useful to set $s_{p+1}=0$.

The symmetry $\Ll_i$ of irreducible fiber Lorentz tensors is given by
\be\Ll_i=\begin{cases}\Y{s_1-1,...,s_q-1,s_{q+1},...,s_p}, & i=0,\\
\Y{s_1-1,...,s_{q-1}-1,s_q-t,s_{q+1},...,s_p}, & i=1,\\
\Y{s_1-1,...,s_{q-i}-1,s_{q-i+2}-1,...,s_q-1,s_q-t,s_{q+1},...,s_p}, & i=2,...,q-1,\\
\Y{s_2-1,...,s_q-1,s_q-t,s_{q+1},...,s_p}, & i=q.
\end{cases}\ee The grade $g$ of $\Ll_0$ is $(s_1-s_q+t-1)$, the grade of $\Ll_i$ is $(s_1-s_{q-i+1})$.
For instance, the physical field $\phi^{\Ss_0}$ is embedded into the frame field $e^{\Ll_0}\fm{q}$ that is defined as
\begin{align}
e^{{a(s_1-1),...,a_q(s_q-1),b(s_{q+1}),...,u(s_p)}}\fm{q}&=
\boldsymbol{\Pi}\left[W^{{a_1(s_1-1),...,a_q(s_q-1),b(s_{q+1})\bullet(\gamma_{q+1}),...,u(s_p)\bullet(\gamma_p),\bullet(\gamma_{p+1})}}\fm{q}\right],\nonumber
\end{align}
where $\boldsymbol{\Pi}$ is a projector that removes the trace part and makes the fiber tensor be traceless, the Young symmetry
conditions hold by virtue of Lemma-A.

The general rule is that to obtain the Lorentz connection $(q-i)$-form embedded either into $W^\Ds\fm{q}$ for $i=0$ or into
$\xi^\Ds\fm{q-i}$ for $i>0$, which is either the frame field $e^{\Ll_0}\fm{q}$ for $i=0$ or the level-$i$ gauge parameter
$\xi^{\Ll_i}\fm{q-i}$ for $i>0$, the maximal number of compensators is contracted with the rows $(q-i+1),...,(p+1)$, which
guarantees by virtue of Lemma-A that the fiber Lorentz tensor has the symmetry of $\Ll_i$, the projector to the traceless part is
needed though.

\paragraph{Lorentz Frame-like down to Lorentz metric-like.}
The physical field $\phi^{\Ss_0}$, the first level gauge parameter $\xi^{\Ss_1}$, ..., the level-$q$ gauge parameter
$\xi^{\Ss_q}$ are embedded into Lorentz connections $e^{\Ll_0}\fm{q}$, $\xi^{\Ll_1}\fm{q-1}$, ..., $\xi^{\Ll_q}\fm{0}$ as the
highest weight parts. For instance, the physical field $\phi^{\Ss_0}$ is embedded into $e^{\Ll_0}\fm{q}$ as
\begin{align}\label{GCPhysicalFieldEmbedding}
\phi^{a_1(s_1),...,a_q(s_q),b(s_{q+1}),...,u(s_p)}&=
\boldsymbol{\Pi}\left[e^{a_1(s_1-1),...,a_q(s_q),b(s_{q+1}),...,u(s_p)|a_1...a_q}\right],
\end{align}
where $\boldsymbol{\Pi}$ is the projector to the traceless part since by virtue of Lemma-B the tensor in brackets has the
symmetry of $\Ss$ but is not traceless. Certain nontrivial traces, which are present in $e^{\Ll_0}\fm{q}$, $\xi^{\Ll_1}\fm{q-1}$,
..., $\xi^{\Ll_q}\fm{0}$, are necessary for the gauge symmetry to be realized off-shell without making gauge parameters be
subjected to (\ref{FullSystemB})-like constraints.

\paragraph{Equations of motion.} Let us now discuss the equations of motion that after imposing certain gauge
lead to (\ref{FullSystemA})-(\ref{FullSystemE}) with the correct mass-like term determined by $(\Ss,q,t)$.

First, note that imposing $R^\Ds\fm{q+1}=\DO W^\Ds\fm{q}=0$ leads to too strong conditions. Actually, $\DO W^\Ds\fm{q}=0$ can be
treated \cite{Bekaert:2005vh} as a sort of cocycle condition, having only pure gauge solutions $W^\Ds\fm{q}=\DO \xi^\Ds\fm{q-1}$
unless $q=0$ by virtue of the Poincare Lemma.

For example, a massless spin-two field, i.e. the gravity linearized over $\dSAdS$, can be described by a single one-form
connection $W^{A,B}_\mu dx^\mu$ of the (anti)-de Sitter algebra, which gives rise to the dynamical\footnote{This can be somewhat
confusing because $\Omega^{A,B}$ describes the background geometry. $W^{A,B}\fm{1}$ describes the small fluctuations over the
(anti)-de Sitter background.} frame $e^a_\mu dx^\mu$ and to the dynamical connection $\omega^{a,b}_\mu dx^\mu$. The field
strength $R^{A,B}\fm{2}=\DO W^{A,B}\fm{1}$ consists of two Lorentz components $R^a\fm{2}=\DL e^a\fm{1}-\lambda
h_m\omega^{a,m}\fm{1}$ and $R^{a,b}\fm{2}=\DL\omega^{a,b}\fm{1}+\lambda h^ae^b\fm{1}-\lambda h^be^a\fm{1}$, which are the
linearized torsion and the Riemann curvature two-form, respectively. By virtue of $R^{a}\fm{2}=0$, $\omega^{a,b}\fm{1}$ is
expressed in terms of the first derivative of $e^{a}\fm{1}$. The dynamical second order equations results from \be
h^{a\mu}h^\nu_cR^{a,c}_{\mu\nu}=0\label{BadEquations}.\ee Obviously, setting the whole field strength $R^{a,b}\fm{2}$ to zero do
not describe any propagating degrees of freedom. Instead of using the operations beyond the class of differential forms as in
(\ref{BadEquations}), we can parameterize the components of the field strength that are allowed to be nonzero on-mass-shell by
the Weyl tensor $C^{aa,bb}\fm{0}$ that is an irreducible Lorentz tensor having the symmetry of $\smallpic{\YoungBB}$. Then,
(\ref{BadEquations}) is equivalent to $R^{a,b}\fm{2}=h_mh_nC^{am,bn}\fm{0},$ or, in manifestly (anti)-de Sitter covariant terms,
to \be R^{A,B}\fm{2}=E_CE_DC^{AC,BD}\fm{0},\ee where $C^{AB,CD}\fm{0}$ is an irreducible tensor of the (anti)-de Sitter algebra
having the symmetry of $\smallboldpic{\YoungBB}$ and it is orthogonal to $V^C$.

Turning back to the general case, the operator \Sigm{} in (\ref{GCLorentzGaugeLawIrr}) accounts for the algebraic and
differential relations between the fields, the gauge parameters and the field strengths. The representatives of the
\Sigm-cohomology groups correspond \cite{Lopatin:1987hz, Shaynkman:2000ts, Skvortsov:2009} to dynamical fields (the field is
called dynamical if it cannot be gauged away by some Stueckelberg symmetry and it is not expressed in terms of derivatives of
other fields), to differential gauge parameters (which cannot be set to zero by higher level Stueckelberg gauge symmetry and are
not Stueckelberg parameters for some fields) and to independent gauge invariant equations. In \cite{Skvortsov:2009} the
\Sigm-cohomology groups are calculated and it is shown that $\phi^{\Ss_0}$, $\xi^{\Ss_1}$, ..., $\xi^{\Ss_q}$ and the equations
discussed below are the representatives of the \Sigm-cohomology groups.

Indeed, it is obvious that the gauge parameters $\xi^{..}\fm{q-1}$ contributing as $\Sigm(\xi^{..}\fm{q-1})$ to the gauge
transformations (\ref{GCLorentzGaugeLawIrr}) for $e^{\Ll_0}\fm{q}$ do not contain in the decomposition into irreducible tensors
of the Lorentz algebra the component with the symmetry of $\Ss_0$, which might be used to gauge away $\phi^{\Ss_0}$. Obviously,
the same holds for $\xi^{\Ss_i}$. Note that certain traces needed for an off-shell description belong to the \Sigm-cohomology
too, these are more hard to find \cite{Skvortsov:2009}.

Due to the Bianchi identity $\DO R^\Ds\fm{q+1}=0$, most of the components of the field strength either can be set to zero by a
nonsingular algebraic field redefinition or are expressed in terms of derivatives of other components. The analysis of the
$\Sigm$-cohomology \cite{Skvortsov:2009} directly implies that the independent gauge invariant differential equations on
$\phi^\Ss$ are given by (1) certain components of the torsion-like field strength $R^{\Ll_0}\fm{q+1}$, which are the first order
differential equations that after fixing certain gauge reduce to (\ref{FullSystemB}); (2) certain components of the field
strengths $R^{\Ll^i_0}\fm{q+1}$ that have one fiber index more as compared to $\Ll_0$ and among which is the component with same
symmetry $\Ss$ as the dynamical field $\phi^\Ss$, which after fixing certain gauge reduces to the wave equation
(\ref{FullSystemA}); (3) the generalized Weyl tensor that is an irreducible Lorentz tensor with the symmetry of \be
\Y{s_1,...,s_q,s_q-t+1,s_{q+2},...,s_p},\ee embedded into the field strength $R^{\Ll_{-1}}\fm{q+1}$ with
$\Ll_{-1}=\Y{s_1-1,...,s_q-1,s_q-t,s_{q+2},...,s_p}$.

What we prove is that the wave equation for $\phi^\Ss$ has the correct mass-like term and, hence, an $(\Ss,q,t)$ gauge field can
indeed be described by the single connection $W^\Ds\fm{q}$ of the (anti)-de Sitter algebra. The technicalities concerning the
\Sigm-cohomology are in \cite{Skvortsov:2009}.

\paragraph{Towards unfolded equations.} In order for the module $\Irrep{E_0}{\Ss}$ with $E_0$ determined by $(\Ss,q,t)$ to be realized on the solutions of equations
of motion one must set to zero all components of the field strength except for the Weyl tensor together with all components of
the field strength that are expressed in terms of its derivatives, these can be embedded into the irreducible tensor
$C^\Ww\fm{0}$ of the (anti)-de Sitter algebra having the symmetry of $\Ww=\Y{s_1,...,s_q-t+1,s_{q+1},...,s_p}$ and satisfying
certain $V$-conditions, so that the equations read\be
R^{A(s_1-1),...,B(s_q-1),C(s_q-t),D(s_{q+1}),...,F(s_p)}\fm{q+1}=E_{L}...E_ME_{N}C^{A(s_1-1)L,...,B(s_q-1)M,C(s_q-t)N,D(s_{q+1}),...,F(s_p)}\fm{0}\ee
The Bianchi identities for the field strength imply that $\DL C^\Ww\fm{0}$ cannot be arbitrary. In the spirit of the unfolded
approach the constraints on $\DL C^\Ww\fm{0}$ can be solved in terms of some other field $C^{\Ww_1}\fm{0}$, for which $\DL
C^{\Ww_1}\fm{0}$ is also constrained and so on. The Weyl tensor together with its descendants form certain infinite-dimensional
module $C$ of the (anti)-de Sitter algebra. The unfolded equations should read
\begin{align*} \DO W^\Ds\fm{q}&=E...E C^{\Ww}\fm{0}, \\ \DOTwisted C^{\Ww_i}\fm{0}&=0, \end{align*}
where $\DOTwisted$ is the $\dSAdS$-covariant derivative in the Weyl module. Note that $\DO$ acts by the adjoint action and
$\DOTwisted$ acts by the twisted-adjoint action in the well-known case of massless spin-$s$ fields  \cite{Vasiliev:1989yr,
Vasiliev:1990en, Vasiliev:2001wa, Vasiliev:2003ev}.

The full unfolded equations for massless fields of the series $(\Ss,q_{min},1)$, where $q_{min}$ is the number of the first equal
rows of $\Ss$, were constructed in \cite{Boulanger:2008up, Boulanger:2008kw}. The explicit realization for $\DOTwisted$ was
obtained for all $(\Ss,q,1)$ fields. More precisely, in \cite{Boulanger:2008up, Boulanger:2008kw} the unfolded equations for
massless $(\Ss,q_{min},1)$ fields were obtained by taking the limit of critical mass (\ref{MassFormula}) in the unfolded
equations for massive spin-$\Ss$ field in \dSAdS{} which result from the radial reduction of the unfolded equations for massless
spin-$\Ss$ field in Minkowski space found recently in \cite{Skvortsov:2008vs}. We expect the approach of \cite{Boulanger:2008up,
Boulanger:2008kw} can give the unfolded equations for all cases, which remains to be elaborated though.

\paragraph{Mass-like term calculation.} That the dynamical field embedded into $W^\Ds\fm{q}$ is a field of the Lorentz algebra
forces us to single out certain Lorentz components of the field strength in order to verify that the correct equations are indeed
imposed on $\phi^\Ss$. The explicit use of projectors similar to (\ref{GCLorentzGaugeLawIrr}) seems to be very complicated. To
get rid of the projectors in intermediate calculations we work with raw Lorentz tensors that are not generally irreducible. At
the final stage, the component with the symmetry of $\Ss$ is recovered by virtue of Lemmas A-B. All expressions are considered
modulo the terms that do not contribute to the highest weight part of the generalized frame, i.e. to $\phi^\Ss$, since we are
going to recover (\ref{FullSystemA}).

Let us consider the raw field strengths $R$ and $R^k$ for the raw generalized frame field $\te$ and for its associated raw
auxiliary fields $\tomega^k$ that have one fiber index more than $\te$. On the {\it r.h.s.} of the expressions for the field
strengths we ignore\footnote{To avoid working with numerous indices it is worth using oscillators. However, having all indices
written explicitly it is easier to see certain nontrivial consequences of the Young symmetry conditions and to discard the terms
irrelevant for the mass-like term of $\phi^\Ss$. } the groups of indices that coincide with
$a_1(s_1-1),...,a_q(s_q-1),b(s_{q+1})\bullet(\gamma_{q+1}),...,u(s_p)\bullet(\gamma_p),\bullet(\gamma_{p+1})$. The form indices
$\mu_1,...,\mu_{q+1}$ have been converted to the fiber indices $\Sa_1,...,\Sa_{q+1}$ so that  the antisymmetrization over
$\Sa_1,...,\Sa_{q+1}$ is implied. With the signs for the anti-de Sitter case, $R$ and $R^k$ read
\begin{align}
R\equiv R^{a_1(s_1-1),...,a_q(s_q-1),b(s_{q+1}),...,u(s_p)|\Sa[q+1]}\fm{q+1}&=\nonumber\\=D^{\Sa}W^{...|\Sa[q]}\fm{q}+&
\lambda\sum_{k=q+1}^{k=p+1}\gamma_{k}W^{...,c(s_k)\Sa\bullet(\gamma_k-1),...|\Sa[q]}\fm{q}+...,\label{MainFieldStrengthA}
\end{align}\begin{align}
R^k&\equiv R^{a_1(s_1-1),...,a_q(s_q-1),b(s_{q+1});...;c(s_k+1);...;u(s_p)|\Sa[q+1]}\fm{q+1}=
D^{\Sa}W^{...,c(s_k+1)\bullet(\gamma_{k}-1),...|\Sa[q]}\fm{q}+\nonumber\\
&+\lambda\sum_{\substack{i=q+1\\i\neq k}}^{i=p+1}\gamma_{i}
W^{...,c(s_k+1)\bullet(\gamma_k-1),...,f(s_i)\Sa\bullet(\gamma_i-1),...|\Sa[q]}\fm{q}+
\lambda(\gamma_{k}-1)W^{...,c(s_{k}+1)\Sa\bullet(\gamma_{k}-2),...|\Sa[q]}\fm{q}+\nonumber\\
&\qquad-\lambda\sum_{\substack{i=q+1\\i\neq
k}}^{i=p+\delta_{p+1,k}}\eta^{f\Sa}W^{...,c(s_k+1)\bullet(\gamma_k-1),...,f(s_i-1)\bullet(\gamma_i+1),...|\Sa[q]}\fm{q}
-\lambda\eta^{c\Sa}W^{...|\Sa[q]}\fm{q}+\nonumber\\
&\qquad\qquad\qquad-\lambda\sum_{i=1}^{i=q}\eta^{a_i\Sa}W^{...,a_i(s_i-2)\bullet,...,c(s_k+1)\bullet(\gamma_{k}-1),...|\Sa[q]}\fm{q}.\label{MainFieldStrengthB}
\end{align}

Each $R^k$ contains in its decomposition into irreducible Lorentz tensors the component with the symmetry of $\Ss$, which can be
obtained by symmetrizing $\Sa_1$, ..., $\Sa_q$ with $a_1(s_1-1)$, ..., $a_q(s_q-1)$, respectively, and, then, taking the trace
with respect to $\Sa_{q+1}$ and an extra index $c$ in the $k$-th group of symmetric indices. Denoting this projector
$\boldsymbol{\pi}_{i}(R^i)$, one obtains \be\boldsymbol{\pi}_{i}(R^i)=D\mbox{tr}(\tomega^i)-D\cdot\tomega^i+\lambda
M_i\phi^\Ss,\ee where $\mbox{tr}(\tomega^i)$ refers to certain trace with respect one fiber and one form index of $\tomega^i$ and
$D\cdot$ stands for the contraction of the (anti)-de Sitter covariant derivative with certain fiber index of $\tomega^i$. The
mass-like term $M_k$ is equal to \begin{align}
M_k=&\gamma_{q+1}+...+\gamma_{k-1}+\gamma_{k}-1+\gamma_{k+1}+...+\gamma_{p+1}+\nonumber\\
&\underbrace{-1-1....-1}_{q}-(\gamma_{q+1}+1)-...-(\gamma_{k-1}+1)+d+s_k-q\label{MainMassA},
\end{align}
where the terms in the first line results from $\boldsymbol{\sigma_-}$-like terms; each term with $\eta^{\Sa a_i}$, $i=1,...,q$
gives $(-1)$, the terms with $\eta^{\Sa f}$ where $f$ belongs to the groups $i=(q+1),...,(k-1)$ give $-(\gamma_i+1)$; $\eta^{\Sa
c}$ produces $(d+s_k-q)$, where for $k=p+1$ $s_k=0$; the rest of terms brings nothing inasmuch as after taking the trace the
indices appear to be rearranged in the way that has no components with the symmetry of $\Ss$. (\ref{MainMassA}) reduces to
\begin{align}
M_k&=(d-q-k+s_k+s_{k-1}-\delta_{q+1,k}t) & k&\in[q+1,p+1]
\end{align}

Naively, one might consider only one field strength, say $R^k$, however, acting this way nontrivial Young symmetrizers come into
play inevitably when expressing $\tomega^k$ from equation $R=0$, (\ref{MainFieldStrengthA}). To simplify calculations we notice
that from $R=0$ one can easily express a linear combination of the form $\sum_{i=q+1}^{i=p+1}\gamma_{i}\tomega^i$ and, hence, the
simplest way to obtain the wave equation on $\phi^\Ss$ is to compute \be
\sum_{i=q+1}^{i=p+1}\boldsymbol{\pi}_{i}(R^i)\gamma_i.\label{MainStrengthsSum}\ee Note that the field strengths $R$, $R^i$ obey
certain Bianchi identities coming from $\DO R^{\Ds}\fm{q+1}=0$. However, (\ref{MainStrengthsSum}) is independent of the Bianchi
identities and can be considered as the dynamical equation for $\phi^\Ss$.

To express $\sum_{i=q+1}^{i=p+1}\gamma_i\tomega^i$ from $R=0$ one symmetrizes $\Sa_1,...,\Sa_q$ with
$a_1(s_1-1)$,...,$a_q(s_q-1)$, to express $\sum_{i=q+1}^{i=p+1}\gamma_i\mbox{tr}(\tomega^i)$ one takes the trace. After little
algebra one obtains the wave equation of the form \be \Box
\phi^\Ss+D(D\cdot\phi^\Ss+D\mbox{tr}(\phi^\Ss))+\lambda^2\sum_{i=q+1}^{i=p+1}\gamma_iM_i\phi^\Ss+[D,D]\te=0.\ee The terms in
brackets can be set to zero by imposing certain gauge. $[D,D]\te$ reads
\be\sum_{i=1}^{i=q}(-)^{i-1}[D^{a_i},D_v]\te^{a_1(s_1-1),...,a_q(s_q-1),b(s_{q+1}),...,u(s_p)|a_1...\hat{a}_i...a_{q}v},\ee and
is equal to $-\lambda^2\sum_{i=1}^{i=q}(d+s_i-q-1)\phi^{\Ss}$ modulo terms that correspond to certain traces. Finally, the total
contribution to the mass-like term in front of $\phi^{\Ss}$ reads \be m^2=\lambda^2\left(\sum_{i=q+1}^{i=p+1}\gamma_i
M_i-\sum_{i=1}^{i=q}(d+s_i-q-1)\right).\ee The direct summation yields (\ref{MassFormula}) as desired. There is no need to do
individual calculations for the gauge parameters inasmuch as having a wave equation for $\phi^\Ss$ with the correct mass-like
term and a gauge parameter with the proper symmetry\footnote{The presence of gauge parameter having the symmetry of $\Ss_1$ that
contributes to the gauge transformations for $\phi^\Ss$ is also important because there might be a formulation without any gauge
symmetry, which describes a 'massive' field.} there are no solutions other than (\ref{ExactSequence}).

On the other hand we can consistently replace $q$ with $(q-i)$, $i=1,...,q$ in
(\ref{MainFieldStrengthA}-\ref{MainFieldStrengthB}). The gauge parameter $\xi^\Ds\fm{q-i}$ will play the role of the gauge field
$W^{\Ds}\fm{q-i}$; the gauge parameters $\xi^\Ds\fm{q-i-1}$, ..., $\xi^\Ds\fm{0}$ will remain to be the gauge parameters for
$\xi^\Ds\fm{q-i}$; the components of the 'field strength' $\xi^\Ds\fm{q-i+1}$ that are to be set to zero plays the role of gauge
fixing conditions.

\paragraph{The results.} Thus, we have shown that the generalized connections of the (anti)-de Sitter algebra
$W^\Ds\fm{q}$, $\xi^\Ds\fm{q-1}$, ..., $\xi^\Ds\fm{0}$ contain as the Lorentz components the dynamical field $\phi^{\Ss_0}$, the
first level gauge parameter $\xi^{\Ss_1}$, ..., the $q$-th level gauge parameter $\xi^{\Ss_q}$ and setting certain components of
the field strength $R^\Ds\fm{q+1}$ to zero we derive for $\phi^\Ss$ the wave equation with the correct mass-like term
(\ref{MassFormula}). Consequently, the exact sequence (\ref{ExactSequence}) is embedded into (\ref{MainGaugeTransformations}).

Since certain gauge connection $W^{\Ds_{\Ss,q,t}}\fm{q}$ is associated with each triple $(\Ss,q,t)$, the map $\varrho$ from the
variety of $\dSAdS$ gauge fields to the variety of $\dSAdS$ connections is an into mapping. Despite the fact that $\varrho$ is
not an onto mapping the rest of gauge connections does not describe anything new, providing us with dual formulations.

First of all, there are three Hodge-like dualities: (1) with the help of the world Levi-Civita symbol $\epsilon_{\mu_1...\mu_d}$
a degree-$q$ form can be transformed to a degree-$(d-q)$ one; (2) the invariant tensor $\epsilon_{A_1...A_{d+1}}$ of the anti-de
Sitter algebra $\mathfrak{g}$ allows to consider the tensors of $\mathfrak{g}$ having the symmetry of a Young diagram with at
most $[(d+1)/2]$ rows, as we have done throughout this paper; (3) on-mass-shell, one can use the invariant tensor
$\epsilon_{\aI_1...\aI_{d-1}}$ of the '(anti)-de Sitter Wigner little algebra' \msv{} to map physical polarization tensors having
the symmetry of a height-$k$ Young diagram into the tensors having the symmetry of a height-$(d-1-k)$ Young diagram.

By the construction, the degree $q$ is constrained by $1\leq q\leq q_{\max}=[(d-1)/2]$ \footnote{$q$ gets its maximal value iff
$q=p$ and $p$ is equal to the maximal height allowed for Young diagrams of \msv, i.e. is equal to $[(d-1)/2]$.}. The forms of
degree zero are not gauge inasmuch as there is no degree-$(-1)$ forms to become gauge parameters. Zero forms play an important
role within the unfolded approach, forming the Weyl module that carries physical degrees of freedom.  Making use of duality-(1)
allows one to map forms of degree higher than $q_{\max}$ into the forms of degree not greater than $q_{\max}$ except for the gap
for $d=2n$ in degree-$n$ forms. Indeed, in this case $q_{\max}=n-1=[(2n-1)/2]$ and, hence, degree-$n$ forms can be obtained from
our construction neither directly nor by means of the duality-(2)\footnote{Having a degree-$n$ form for $d=2n$ suggests imposing
(anti)-self duality conditions with respect to world form indices \cite{Metsaev:2008ba}, which seems problematic since the Hodge
operator is built of the metric field that is to become a dynamical field in the full interacting theory.}. However, no new
\dSAdS{} gauge fields arise in this way inasmuch as a height-$n$ diagram of $\msv$ is equivalent to the height-$(n-1)$ diagram by
means of the duality-(3). Consequently, there are two equivalent formulations for any gauge field $(\Ss,q,t)$ in $\dSAdS$ except
for $d$ even and $q=q_{\max}$, in which case there are three equivalent formulations.

\section{Discussion and Conclusions}\label{SConclusions}\setcounter{equation}{0}

Each $\dSAdS$ gauge field is uniquely defined by a triple $(\Ss,q,t)$ \cite{Alkalaev:2009} consisting of an $\msv$ Young diagram
that characterizes spin degrees of freedom and integer parameters $q$, $t$ that determine the gauge symmetry - the gauge
parameter has the symmetry of a diagram obtained by removing $t$ cells from the $q$-th row of $\Ss$, the order of derivatives in
the gauge transformation law is equal to $t$.

We have shown that the gauge connection $W^{\Ds_{\Ss,q,t}}\fm{q}$ with values in the irreducible module of the (anti)-de Sitter
algebra $\Ds_{\Ss,q,t}$ defined by (\ref{GFGCAdSDiagram}), is a natural geometric framework for gauge field $(\Ss,q,t)$. The
whole set of auxiliary fields is incorporated into the single $q$-form. The gauge transformations have very simple form and the
field strength is manifestly gauge invariant.

The frame-like Lorentz formulation is obtained by performing the dimensional reduction of the tensor $\Ds$ of the (anti)-de
Sitter algebra down to irreducible tensors of the Lorentz algebra. The metric-like formulation is obtained by further decomposing
the connection of the Lorentz algebra into fully metric-like tensors.

As soon as we have identified the free field theory described by the connection $W^\Ds\fm{q}$ there is no need in decomposing the
$\dSAdS$ module $\Ds$ into the Lorentz ones, taking the advantage of working in terms of a single field that has a clear
algebraic and geometric meaning.

Notwithstanding the fact that only bosonic fields were considered in this paper, the extension to the fermionic fields is
straightforward, more complicated though due to Majorana, Weyl, and Majorana-Weyl conditions to be analyzed carefully. We
conjecture the final conclusion to be the same in that a fermionic gauge field defined by $(\Ss,q,t)$, where $\Ss$ refers to the
tensor part of an irreducible $\msv$-spin-tensor, can be described by a gauge connection $W^{\hat{\alpha}:\Ds_{\Ss,q,t}}\fm{q}$,
where the tensor part is obtained by the same rules as in the bosonic case and $\hat{\alpha}$ is a spinor index of the
$\dSAdS$-algebra.

A number of dual formulations is also included in $W^{\Ds_{\Ss,q,t}}\fm{q}$ - any of the auxiliary fields at grade higher than
that of the frame field can be regarded as a dynamical one inasmuch as by setting certain components of the field strength to
zero, lower grade fields can be expressed in terms of derivatives of the fields at higher grade. This issue is far beyond the
scope of the paper. As an example, for a massless spin-$s$ field, i.e. $\Ss=\Y{s}$, $q=1$, $t=1$, instead of the frame field
$e^{a(s-1)}\fm{1}$ the auxiliary field at grade-one $\omega^{a(s-1),b}\fm{1}$ was taken to be the dynamical field in
\cite{Matveev:2004ac}.

There is no room for massive fields in this picture since massive fields are nongauge by nature. The potentials for massive
fields are forms of degree zero and belong to certain infinite-dimensional modules of the $\dSAdS$-algebra realized on zero
forms, see \cite{Shaynkman:2000ts, Boulanger:2008up, Boulanger:2008kw}.

We would like to stress that the proposed frame-like description of arbitrary-spin (partially)-massless fields tells us not so
much about the Lagrangian description for the general case, since most of the transversality constraints (\ref{FullSystemB})
cannot be obtained by gauge fixing and, hence, supplementary fields may be needed.

That each $W^{\Ds}\fm{q}$ may describe certain gauge field does not imply that $W^{\Ds}\fm{q}$ cannot be used another way. For
instance, $W^{A,B}\fm{1}$ can be used to describe either a dynamical spin-two field or the background (anti)-de Sitter space
($\Omega^{A,B}$). The theory is defined by the equations imposed in terms of $R^{\Ds}\fm{q+1}$. There exists a powerful method,
known as $\boldsymbol{\sigma_-}$-cohomology \cite{Lopatin:1987hz, Shaynkman:2000ts, Skvortsov:2009}, to classify all gauge
invariant differential equations that can be imposed on $W^{\Ds}\fm{q}$. The matter being very technical,
$\boldsymbol{\sigma_-}$-cohomology are found in a companion paper \cite{Skvortsov:2009}.

To be clarified is the Minkowski limit of the proposed $\dSAdS$ systems. The Poincare algebra has no tensor representations,
hence, to take the Minkowski limit $W^{\Ds}\fm{q}$ has to be reduced to the connections of the Lorentz algebra. The Minkowski
limit of a massless or partially-massless field is given generally by a direct sum of Minkowski massless fields
\cite{Brink:2000ag}. Both the frame-like and the unfolded descriptions of arbitrary-spin massless field in the Minkowski space
are known \cite{Skvortsov:2008vs, Skvortsov:2008sh}. After appropriate rescaling of Lorentz connections $\omega^{\Ll_\alpha}$
arising from $W^{\Ds}\fm{q}$, with the help of the background vielbein $h^a_\mu$ one can construct a map that takes each
$\omega^{\Ll_\alpha}$ to a field from the unfolded system of certain Minkowski massless field that is present in the Minkowski
limit according to \cite{Brink:2000ag}.

Gauge fields in \dSAdS{} are in fact more massive as compared to their Minkowski massless partners inasmuch as only one gauge
symmetry survives in \dSAdS{} and it kills a small part of degrees of freedom. Therefore, one may argue that gauge fields in
\dSAdS{} should be reformulated in much the same way as massive fields \cite{Zinoviev:2002ye, Zinoviev:2001dt, Zinoviev:2003dd,
Zinoviev:2003ix,Zinoviev:2008ze,Zinoviev:2008ve,Zinoviev:2009vy}, with the rest of gauge symmetries that get broken in $\dSAdS$
to be restored upon introducing Stueckelberg fields. Nonminimal Stueckelberg formulation of this sort was constructed in
\cite{Boulanger:2008up, Boulanger:2008kw}. It would also be interesting to track the appearance of all gauge potentials
$W^{\Ds}\fm{q}$ for a fixed $\Ss$ from the results of \cite{Boulanger:2008up, Boulanger:2008kw}.

In the framework of the unfolded approach the connection $W^\Ds\fm{q}$ constitutes the gauge sector of the unfolded system of
equations. Following the success of the unfolded approach for massless spin-$s$ fields, in which case the full nonlinear theory
of one-form connections $W^\Ds\fm{1}$ with $\Ds=\parbox{1.7cm}{\unitlength=0.25mm\boldpic{\RectBRow{6}{6}{$\scriptstyle
s-1$}{$\scriptstyle s-1$}}}$ was constructed in \cite{Vasiliev:1990en, Vasiliev:2003ev}, we expect the proposed formulation for
gauge fields in $\dSAdS$ to play an important role in the nonlinear theories of mixed-symmetry fields, which are believed to
exist \cite{Metsaev:1993mj, Metsaev:1993ap, Metsaev:2005ar, Boulanger:2004rx}.

\section*{Acknowledgements}
The author is grateful to M.A.Vasiliev, K.B.Alkalaev, O.V.Shaynkman, R.R.Metsaev, N.Boulanger and P.Sundell for many illuminating
discussions on mixed-symmetry fields. The author wishes to thank M.A.Vasiliev for reading the manuscript and giving many valuable
comments. The work was supported in part by grants RFBR No. 08-02-00963, LSS-1615.2008.2, INTAS No. 05-7928, by the Landau
scholarship and by the scholarship of the Dynasty foundation.

\providecommand{\href}[2]{#2}\begingroup\raggedright\endgroup

\end{document}